\documentclass[]{spie}  

\usepackage{amsmath,amsfonts,amssymb}
\usepackage{graphicx}
\usepackage{siunitx}[=v2]
\usepackage[colorlinks=true, allcolors=blue]{hyperref}
\usepackage[table]{xcolor}
\usepackage{diagbox}
\usepackage{multirow}
\usepackage{todonotes}
\usepackage{float}
\usepackage{wasysym}

\title{Additive manufacturing in aluminium of a primary mirror for a CubeSat application: manufacture, testing and evaluation.}

\author[a]{Ilhan Aziz}  
\author[a]{Younes Chahid}
\author[a]{Jennifer Keogh}
\author[a]{James Carruthers}
\author[a]{Katherine Morris}
\author[a]{Joel Harman}
\author[a]{Scott McPhee} 
\author[a]{Eilidh Fraser}
\author[a]{Luca Millan}
\author[b]{Cyril Bourgenot}
\author[b]{Paul White}
\author[b]{Spencer Davies}
\author[c]{Franck P. Vidal}
\author[d]{Wenjuan Sun}
\author[d]{Mirko Sinico}
\author[e]{Fraser Laidlaw}
\author[f]{Wai Jue Tan}
\author[f]{Arindam Majhi}
\author[a]{Carolyn Atkins}
\affil[a]{UK Astronomy Technology Centre, Royal Observatory, Edinburgh, EH9 3HJ, UK}
\affil[b]{Durham University, NETPark Research Institute, Sedgefield, TS21 3FD, UK}
\affil[c]{UKRI-STFC Scientific Computing, Daresbury Laboratory, Warrington WA4 4AD, UK}
\affil[d]{Department of Mechanical Engineering, KU Leuven, 3001 Leuven, BE}
\affil[e]{School of Physics and Astronomy, University of Edinburgh, EH9 3BF, UK}
\affil[f]{Diamond Light Source, Harwell Campus, Didcot OX11 0DE, UK}
\authorinfo{Further author information: \\E-mail: carolyn.atkins@stfc.ac.uk}

\begin{document} 
\maketitle

\begin{abstract}
Additive manufacturing (AM; 3D Printing), a process which creates a part layer-by-layer, has the potential to improve upon conventional lightweight mirror manufacturing techniques, including subtractive (milling), formative (casting) and fabricative (bonding) manufacturing. Increased mass reduction whilst maintaining mechanical performance can be achieved through the creation of intricate lattice geometries, which are impossible to manufacture conventionally. Further, part consolidation can be introduced to reduce the number of interfaces and thereby points of failure. AM design optimisation using computational tools has been extensively covered in existing literature. However, additional research, specifically evaluation of the optical surface, is required to qualify these results before these advantages can be realised. This paper outlines the development and metrology of an AM mirror for a CubeSat platform with a targeted mass reduction of 60\% compared to an equivalent solid body. This project aims to incorporate recent developments in AM mirror design, with a focus on manufacture, testing and evaluation. This is achieved through a simplified design process of a Cassegrain telescope primary mirror mounted within a 3U CubeSat chassis. The mirror geometry is annular with an external diameter of \SI{84}{\milli\meter} and an internal diameter of \SI{32}{\milli\meter}; the optical prescription is flat for ease of manufacture. Prototypes were printed in AlSi10Mg, a low-cost aluminium alloy commonly used in metal additive manufacturing. They were then machined and single-point diamond turned to achieve a reflective surface. Both quantitative and qualitative evaluations of the optical surface were conducted to assess the effect of hot isostatic pressing (HIP) on the optical surface quality. The results indicated that HIP reduced surface porosity; however, it also increased surface roughness and, consequently, optical scatter.     
\end{abstract}


\section{INTRODUCTION}
\label{sec:intro}  

Additive manufacturing (AM; 3D Printing) provides substantial advantages compared to traditional manufacturing techniques\cite{attaran2017rise}\cite{perez2020AMadvantages}. AM allows for the direct creation of complex geometries without the need for tooling or molds\cite{gibson2021additive}. This capability enables the production of parts with intricate internal structures, such as lattice geometries, which are impossible to achieve using conventional methods. The ability to design and manufacture lightweight components with optimised complex geometries can result in substantial mass reduction whilst retaining stiffness. Furthermore, AM facilitates part consolidation, reducing the number of interfaces, which can enhance the overall reliability of the final design. AM is increasingly being adopted within the aerospace and medical sectors\cite{vranic2017advantages}, enabling design improvements previously thought impossible. Example applications include the Twin Annular Premixing Swirler (TAPS)\cite{foust2012development} combustor for jet engines, which reduces nitrous oxide emissions by 60\% compared to the previous generation of designs through the use of complex air channels, which can only be manufactured using AM. AM has also facilitated custom lighter prosthetics \cite{AMProsthesisReview}, which have increased user comfort. However, the advantages of AM design are not limited to these sectors, with astronomical instrumentation seen as another promising field for AM application \cite{chahid2024additive}.

Telescopes and astronomical instrumentation play a crucial role in advancing our understanding of the universe, enabling detailed observation and analysis of celestial objects and phenomena. Instruments such as cameras and spectrometers are designed to focus or split light across a range of wavelengths, from radio waves to X-rays, providing valuable data for ground- and space-based astronomy. A key component of telescopes and astronomical instrumentation are mirrors, which focus and direct light. The optical performance of these mirrors directly influences the quality and precision of astronomical observations. In recent years, the development of lightweight and high-performance mirrors has become particularly important as the demand for more compact and efficient instruments grows, especially for space-based missions, due to the high, although decreasing, launch costs and limited payload volume of existing launch vehicles\cite{LaunchVehiclesCost}. It is this application need that has driven research in fabricating lightweight (low-mass) mirrors using AM, where the design freedom of AM can be leveraged for the given satellite requirements.

\begin{figure}[htbp]
\begin{center}
\includegraphics[height=7cm]{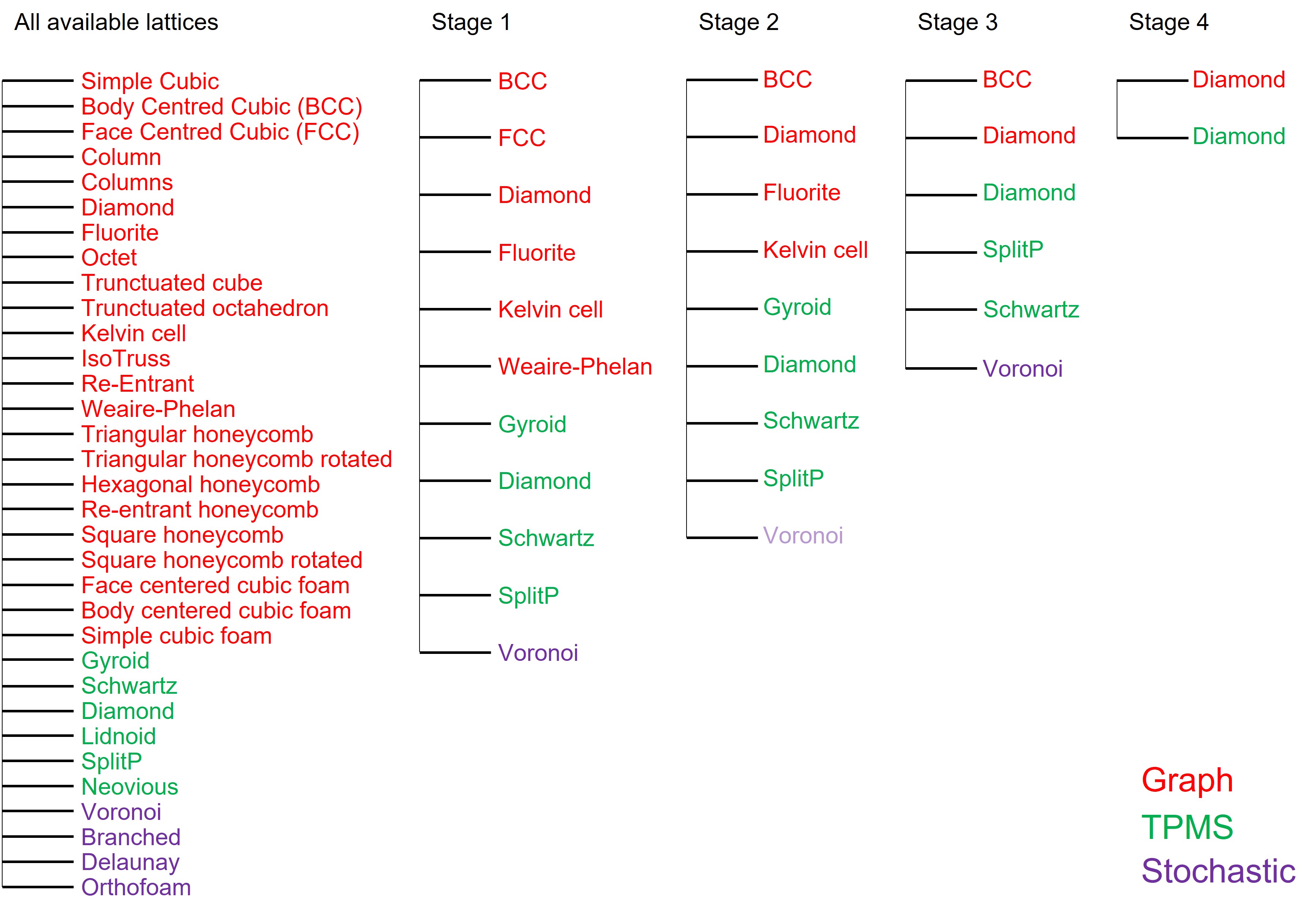}
\end{center}
\caption{Lattice down selection process for a rectangular AM mirror petal. Image credit: \textit{Westik et al. (2022)} \cite{westsik2023design}.}
\label{fig:lattice downselection example}
\end{figure}

The first AM mirror designs focussed on graph (strut-based) lattices \cite{zhang2022review, chahid2024additive}; however, with increased computing capability, the triply periodic minimal surface (TPMS) family of lattices have been shown to have superior strength to weight characteristics as outlined by \textit{Westik et al. (2023)} \cite{westsik2023design}, achieving a mass reduction of 44\%, with an example lattice down-selection shown in Figure~\ref{fig:lattice downselection example}. This design was optimised to minimise the impact of print through effects during polishing, which occur due to unsupported regions within a design deflecting more than the surrounding supported regions, resulting in a mid-spatial frequency form error generated by the lattice as described by \textit{Atkins et al. (2019)} \cite{atkins2019}. \textit{Lister et al. (2024)} \cite{lister2024design} built on these results, investigating further TPMS lattice types, increasing the mass reduction to 70\%. Numerous case studies have been carried out, exploring and optimising lightweight aluminium AM mirrors over the past decade \cite{zhang2022review}. 

Lattice structures can be optimised for their specific design requirements and loading conditions, the most commonly used approach amongst existing AM mirror designs is a topology optimisation \cite{zhang2022review}. A topology optimisation utilises the results from a finite element analysis (FEA) to identify areas of high and low mechanical stress. Material can then be removed accordingly, maximising the strength-to-weight ratio of a part. One such example is that of the design study by \textit{Fan et al. (2021)}\cite{fan2021research},  optimising for minimal structural flexibility, the final design achieved a mass reduction of 81.2\% compared to a solid body equivalent.

 Going from an optimised AM design to a physical part requires selecting an appropriate AM fabrication technology and material, \textit{Chahid et al. (2024)} \cite{chahid2024additive} reviewed published AM mirrors up to 2024 and identified that laser powder bed fusion (L-PBF) in aluminium alloy was the most common combination. L-PBF is a form of laser sintering in which a thin layer of metal powder is selectively melted to match the cross-section of a design. A new layer of metal powder is then applied on top before the process is repeated until the part is complete and the excess powder is removed. This process is advantageous because thin walls, down to $\sim$\SI{100}{\micro\meter}  can be created as investigated by \textit{Calignano et al. (2022)}\cite{calignano2018manufacturing}, allowing for a broad range of structures; however, to guarantee a successful build on an unknown L-PBF machine, such as when using an external print bureau, a \SIrange{0.8}{1}{\mm} minimum wall thickness is commonly used. Aluminium alloys are frequently the preferred material for lightweight AM mirrors because of their long heritage in conventionally manufactured designs due to beneficial material properties such as low density and low hardness, which makes it easier to machine. \textit{Roulet et al. (2018)} \cite{roulet20183d} conducted contact profilometry on as-printed aluminium alloy mirrors, and the results indicated that the sample had an uneven surface with bumps and a root mean square (RMS) surface roughness value of $\sim$\SI{9}{\micro\meter} using a \SI{0.8}{\milli\meter} cut off filter ($\lambda_{c}$ = \SI{0.8}{\milli\meter}). Further research into the surface roughness of as-printed aluminium alloy components using contact profilometry by \textit{Breen et al. (2022)}\cite{breen2022outgassing} found surface roughness with a RMS value of \SI{12.15}{\micro\meter}, which was reduced to \SI{0.24}{\micro\meter} RMS post machining, in both measurements at $\lambda_{c}$ = \SI{0.8}{\milli\meter}. These results demonstrate the poor optical surface quality of as-printed aluminium alloys; therefore, there is a need for post-processing techniques, such as single-point diamond turning (SPDT), to generate a reflective surface.  

SPDT, a post-processing machining technique, achieves good optical quality of AM substrates. SPDT uses an ultra-precision lathe with a diamond cutting tool to remove fine amounts of material in the micrometre range, whilst the part is rotated about the centre of its optical axis. After cutting, the form error and surface roughness are routinely measured to assess the conformance to the desired optical prescription and the likelihood of optical scattering. The surface form error quantifies the deviation of an optical surface from its perfect-shape equivalent. Incorporating SPDT in the manufacturing process of AM AlSi10Mg alloy mirrors, \textit{Fu et al. (2024)} \cite{fu2024lightweight} achieved a RMS surface form of \SI{44}{\nano\meter} and \SI{28}{\nano\meter} for a primary and secondary mirror with clear aperture diameters of \SI{60}{\milli\meter} and \SI{24}{\milli\meter}, respectively. Also using SPDT with AM mirrors, \textit{Tan et al. (2020)} \cite{tan2020design} achieved a RMS surface form error of  \SI{60}{\nano\meter} for an AlSi10Mg aluminium alloy mirror with a clear aperture of \SI{67}{\milli\meter} by \SI{50}{\milli\meter}. In contrast, surface roughness quantifies the smoothness of an optical surface, good surface roughness values for AM aluminium mirrors are between \SI{3}{\nano\meter} to \SI{4}{\nano\meter} RMS \cite{atkins2019, westsik2023design} with surface roughness $>$ \SI{10}{\nano\meter} RMS considered poor quality \cite{herzog2015optical} \cite{fu2024lightweight}. \textit{Sweeney et al. (2015)} outlined the manufacture of an AM mirror with good surface roughness at $<$ \SI{5}{\nano\meter} incorporating SPDT. These results highlight the potential of SPDT to deliver surface form error and roughness of AM aluminium mirrors for visible and infrared applications. Although significant progress has been made in the manufacture of AM mirrors, there remain several challenges, namely porosity and scratch marks, which hinder the surface quality.

During the L-PBF process, pores can form when the laser fails to melt a region of material, resulting in lack of fusion pores or due to over-melting of regions, resulting in keyhole pores as described by \textit{Snell et al. (2022)}\cite{snell2022towards}. The presence of pores within AM parts, which act as nucleation sites for crack growth\cite{tammas2017influence}, can reduce the mechanical strength of AM parts, as investigated by \textit{Moran et al. (2017)} \cite{moran2022hot}. Furthermore, porosity such as that described by \textit{Fu et al. (2024)} \cite{fu2024lightweight} degrades the optical surface by increasing the surface roughness, and therefore the optical scatter. \textit{Cooper, et al. (2019)} \cite{Cooperetal} quantified pores within an AlSi10Mg AM mirror using X-ray computed tomography (XCT) and described the challenges surrounding porosity identification, specifically thresholding to isolate the pores, \textit{Chahid, et al. (2024)} \cite{Chahid_2024}, built on this research by using a reference pin to calibrate the thresholding value to increase the accuracy of porosity detection. External porosity detection methods have also been used,\textit{Wang et al. (2025)}\cite{wang2025effect} used a scanning electron microscope (SEM) to image surface porosity.  One method of reducing porosity, which is commonly applied to AM mirrors\cite{zhang2022review}, is hot isostatic press (HIP). During HIP a part is subject to high temperatures as well as high pressure for several hours, \textit{Snell et al. (2022)} \cite{snell2022towards} demonstrated that HIP closes up internal pores.  However, HIP can bring its own challenges with \textit{Atkins et al. (2024)} \cite{AtkinsHIP2024} linking HIP to increased surface roughness, \textit{Wang et al. (2025)}\cite{wang2025effect} further investigated the effect of HIP on surface porosity, linking HIP to the formation of silicon precipitates, resulting in increased surface roughness. In addition to removing porosity, a second motivation to use HIP could be to reduce scratches on the optical surface created by SPDT interacting with inclusions.

\textit{Fu et al. (2024)} \cite{fu2024lightweight} observed several scratches on the optical surface of an AlSi10Mg alloy AM mirror. Post SPDT scratches in the form of ring bands were clearly visible, which deteriorated the surface optical quality. Scratch marks formed during SPDT were linked to the presence of inclusions by \textit{Wang et al. (2023)}\cite{wang2023surface} and \textit{Wang et al. (2025)}\cite{wang2025effect}, where hard inclusions on the optical surface were dragged along by the diamond cutting tool cutting into the surface. \textit{Snell et al.(2022)} \cite{snell2022towards} demonstrated the potential of HIP in reducing the formation of scratch marks. Additionally, \textit{Groden et al. (2024)} \cite{GRODEN2024} has shown that contaminants present during the L-PBF print process promote surface cracking and, as a result, degrade the mechanical properties of a part. Where contamination within the print material is difficult to avoid, post-processing techniques such as HIP are necessary to achieve a consistent quality optical surface.

Although AM has significant potential for advancing opto-mechanical design in astronomical instrumentation, several challenges need to be addressed for its successful application.  Firstly, whilst much progress has been made in terms of mechanical design and surface quality, a lack of real-world testing data, particularly regarding space qualification, remains a barrier to the practical application of AM mirrors. As discussed by \textit{Chahid et al. (2024)}\cite{chahid2024additive}, most AM mirrors developed as part of existing literature are of a low technology readiness level TRL, between three and four. Testing within a relevant environment (TRL 5) is crucial for validating the reliability and performance of AM mirrors, marking the next tiers of TRL and the final stages before mission adoption.

This paper outlines the development and optical surface evaluation of an AM mirror for a CubeSat platform with a targeted mass reduction of 60\% compared to a solid body equivalent. To deliver the project within one year, an accelerated design process was used to maximise the time spent on testing and evaluation. The aim of which was to assess the effects of each post-processing step on the surface quality ahead of commencing a study into the impact of environmental testing on AM optical surfaces. The post-processing steps carried out were HIP, SPDT, and an optical coating. XCT was used for internal pore identification and quantification. Additionally, external metrology was performed with a focus on the micro-scale of the mirror, utilising microscopy at three different scales: a large field of view, employing a digital microscope; a medium to small field of view, including topography and composition, which utilised a SEM and interferometer.   

The specifications for the mirror, which are based on the 3U CubeSat design, are outlined in Section~\ref{sec:DesignSpec}. Next, the design process is detailed in Section~\ref{sec:DesignProcess}, including the topology optimisation steps for the lattice structure. Section~\ref{sec:Manufacture} describes the manufacturing steps that were followed. An internal porosity analysis was conducted using XCT scanning, which is explained in Section~\ref{sec:Evaluation - internal, XCT}. Section~\ref{sec:OpticalMetrology} covers external metrology, including an analysis of surface porosity. Finally, a summary and outline of future work are presented in Section~\ref{sec:SummaryandFutureWork}.

\section{Design specifications and constraints}
\label{sec:DesignSpec}
\subsection{Design specification}

\begin{figure}[htbp]
\begin{center}
\includegraphics[height=5cm]{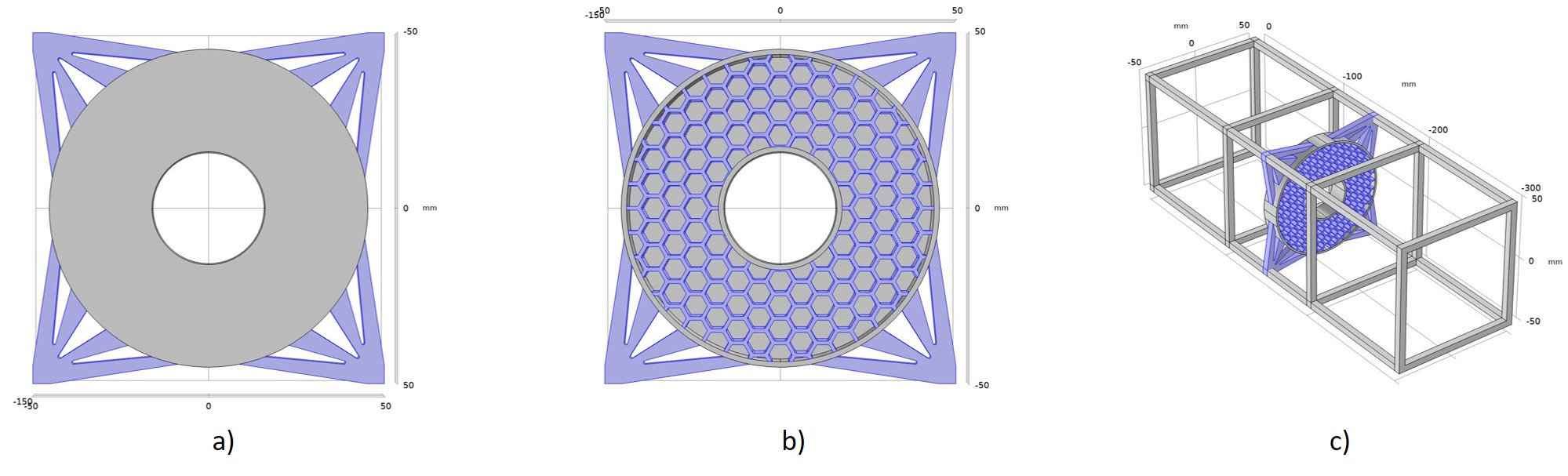}
\end{center}
\caption{A2IM CubeSat mirror early concept, which had its dimensions also taken from the CCam case study: a) front view of the mirror; b) backside of the mirror including the lighweighting structure shown in blue; c) integration within the 3U CubeSat chassis. Image credit: \textit{Snell et al. (2020)} \cite{Snell2020}.}
\label{fig:CCamInfographic}
\end{figure}

The emphasis of the project is on manufacturing, testing, and evaluation, therefore, an existing mirror design was selected as the foundation for the design process. The selected mirror was the primary mirror (M1) from CubeSat Camera (CCam), a low-cost imaging system case study featuring a Cassegrain telescope design, developed by RAL Space and UK ATC \cite{CCamPaper}. This project builds upon a previous CCam AM design concept\cite{Snell2020}, which integrated mass reduction and mounting; a concept design is shown in Figure~\ref{fig:CCamInfographic}.

Except for the optical prescription, all other parameters were derived from the dimensions of the CCam M1 primary annular mirror, with the specifications of the AM mirror shown in Table~\ref{tab:MirrorDimensions}. The mechanical aperture of the AM mirror has a diameter of \SI{84}{\milli\meter}, with a clear aperture of \SI{80}{\milli\meter}. Two walls were incorporated within the AM mirror, which defined the boundaries for the internal lattice similar to a conventional open back design as shown in Figure~\ref{fig:CCamInfographic} b): one along the internal perimeter and another along the external perimeter, each of \SI{1}{\milli\meter} thickness. A flat optical prescription was adopted to simplify the manufacturing and metrology processes; the height of the part was set at \SI{10}{\milli\meter}, with a \SI{1}{\milli\meter} thick mirror surface. Integrated with the mirror is a mounting structure which serves only as a connecting interface to the CubeSat chassis; in this study, an optimum optomechanical design was not required.

\begin{table}[htbp]
\caption{AM prototype mirror specifications.} 
\label{tab:MirrorDimensions}
\begin{center}       
\begin{tabular}{|l|l|} 
\hline
\rule[-1ex]{0pt}{3.5ex}  Design aspect & Specification  \\
\hline
\rule[-1ex]{0pt}{3.5ex}  Mechanical aperture & \SI{84}{\milli\meter}   \\

\rule[-1ex]{0pt}{3.5ex}  Clear aperture & \SI{80}{\milli\meter}   \\

\rule[-1ex]{0pt}{3.5ex}  Wall thickness & \SI{1}{\milli\meter}  \\

\rule[-1ex]{0pt}{3.5ex}  Height & \SI{10}{\milli\meter}   \\

\rule[-1ex]{0pt}{3.5ex}  Optical prescription & Flat  \\

\rule[-1ex]{0pt}{3.5ex}  Mass reduction target & 60\% compared to a solid \\ 
\rule[-1ex]{0pt}{3.5ex}  & body equivalent  \\

\rule[-1ex]{0pt}{3.5ex}  AM method & L-PBF  \\

\rule[-1ex]{0pt}{3.5ex}  AM material & AlSi10Mg aluminium alloy  \\

\rule[-1ex]{0pt}{3.5ex}  Print orientation & Exposed lattice face  \\
\rule[-1ex]{0pt}{3.5ex}  & facing upwards  \\

\rule[-1ex]{0pt}{3.5ex}  Optical fabrication & SPDT  \\

\rule[-1ex]{0pt}{3.5ex}  Target surface roughness & $<$ \SI{10}{\nano\meter} RMS \\

\hline
\end{tabular}
\end{center}
\end{table} 

\begin{figure}[htbp]
\begin{center}
\includegraphics[width = 0.95\textwidth]{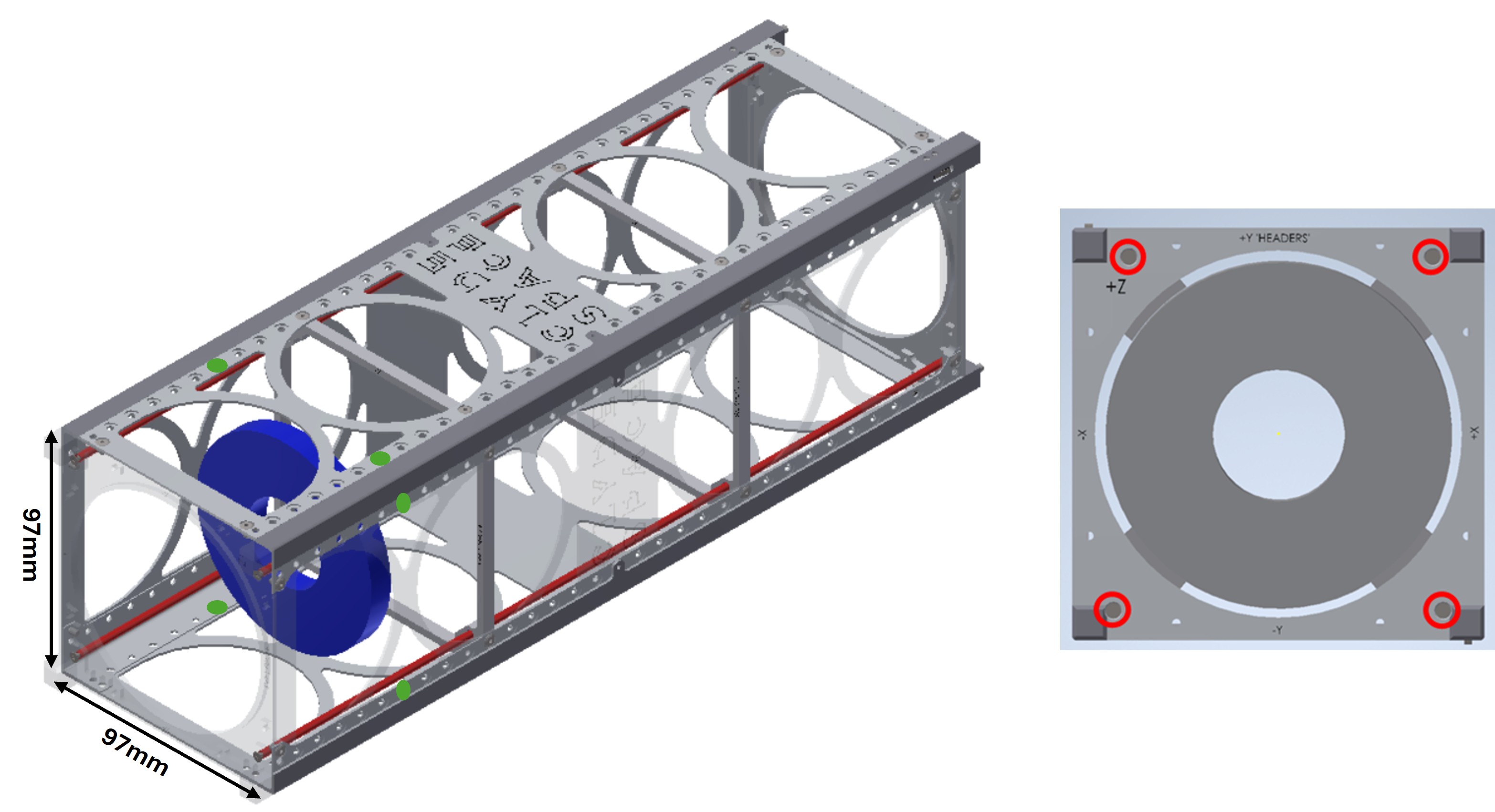}
\end{center}
\caption{Left, AM mirror (blue) positioned within the 3U CubeSat chassis, the mounting holes visible from this perspective are circled in green, and the PCB mounting rods are shown in red; right, asymmetric positioning of the PCB mounting rods, which are circled in red.}
\label{fig:AMmirrorchassisintegration}
\end{figure}

The mounting structure features eight fixing holes, two per corner, located in perpendicular planes as shown in Figure~\ref{fig:AMmirrorchassisintegration}. These eight fixing holes formed a square perimeter of \SI{97}{\milli\meter} by \SI{97}{\milli\meter} bounded by exterior panels. The 3U CubeSat chassis for integration also featured four parallel printed circuit board (PCB) mounting rods that ran the length of the chassis, also shown in Figure~\ref{fig:AMmirrorchassisintegration}. Therefore, clearance holes within the mounting structure were required to accommodate these rods.

\subsection{Manufacturing design constraints}
\begin{figure}[htbp]
\begin{center}
\includegraphics[width = 0.5\textwidth]{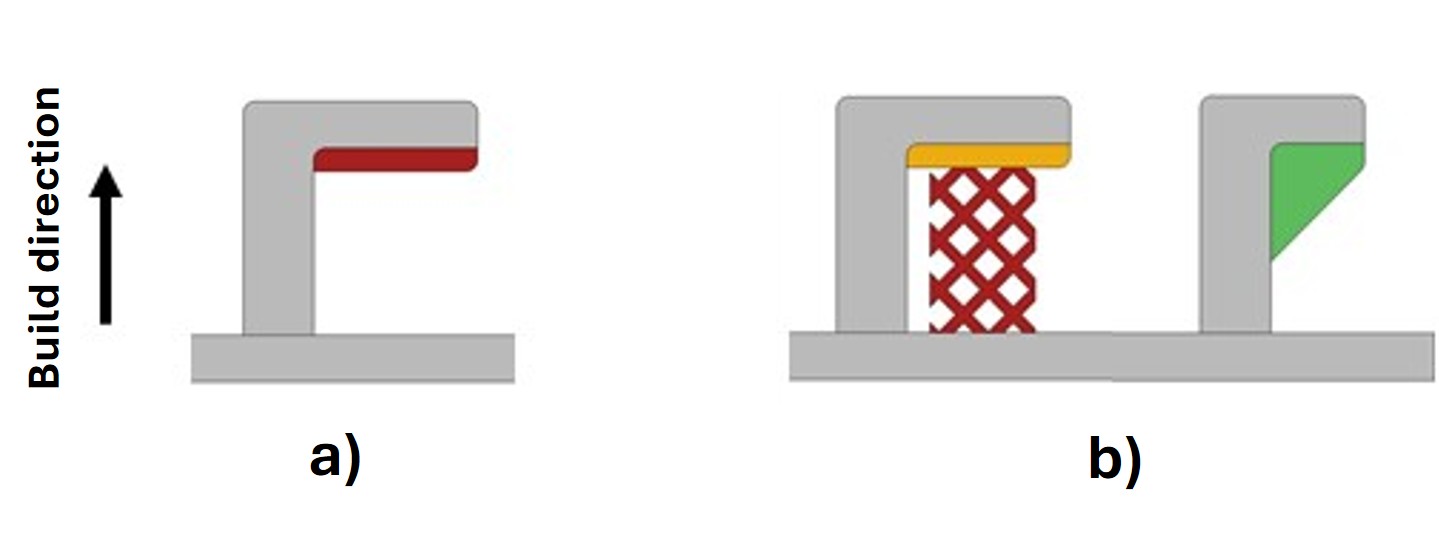} 
\end{center}
\caption{Design which includes an overhang: a) starting design with overhang highlighted in red; b) support (left) and decreased overhang angle (right) design solutions. Image credit: \textit{Atkins et al. (2021)} \cite{AMCookBook}.}
\label{fig:overhang}
\end{figure}

AM offers significant advantages in terms of lightweighting and part consolidation, but it also introduces unique design limitations such as sharp corners, overhangs, minimum feature size, powder removal, and print-through effects. Sharp corners are stress concentrators which cannot be accurately produced by L-PBF and so need to be replaced with fillets or chamfers.  Overhangs (unsupported regions), shown in Figure~\ref{fig:overhang}, are limited by variables such as the angle relative to the build plate, the size of the overhang, and the print settings used. Diverging from the preferred limits on these variables can result in partial or full collapse of the overhang. This can be avoided by changing the print orientation, which needs to be set so that minimal, if any, overhangs are present, or by reducing the overhang angle. Where overhangs cannot be avoided, support material can be introduced. However, this is not an ideal solution, due to the increased associated costs and print times but also because support material within lattice structures cannot be removed. A summary of industry standard L-PBF guidelines to minimise overhangs is presented in \textit{Atkins et al. (2021)} \cite{AMCookBook}, such as a maximum overhang angle of \SI{45}{\degree}. Minimal feature size defines the minimum dimensions for a feature to be accurately produced. It has been shown that the minimum dimensions a feature can have to be accurately produced by L-BPF are \SI{0,50}{\milli\meter} for lattice structures and \SI{0,75}{\milli\meter} for straight wall structures \cite{LPBFMinimumPrintSizeReference}, however, these values vary depending on the machine used as well as the print settings.  Powder removal represents another challenge, particularly for parts produced with L-PBF using lattice structures, such as this study. If the design does not include open back holes, cavities must be carefully designed and optimised to ensure that all excess powder can be effectively removed post-printing. Lastly, print-through effects need to be considered during the design process, to minimise the influence of lattice structures imprinting onto the optical surface due to the non uniform support.

Achieving the dimensional accuracy and flatness required for optical design necessitates post-processing techniques. Therefore, two design drawings were required, an as-printed design as well as an as-machined part, with the differences between the two driven by tolerance and finishing requirements. Furthermore, the mounting structure needed to accommodate the interface to the SPDT fixture plate, so the mounting structure required a fixing interface parallel to the mirror surface. This interface had freedom in terms of its positioning within the square perimeter, due to the use of a fixture plate as a common interface between the AM mirror and the lathe. Also required for SPDT were flat edges to butt the part against, locking rotation in the mirror surface plane. To minimise the risk of AM defects from contamination, AM suppliers using machines with a known history of single material use (AlSi10Mg) were considered in this study.

\subsection{Design requirements}

Based on the design specifications and constraints, the following requirements were set:
\begin{enumerate}
    \item Mounting structure - to accommodate interfacing to a 3U CubeSat chassis incorporating the four PCB rods and to the SPDT fixture with required mechanical constraints in rotation and translation.
    \item Sharp corners - internal and external mirror surface corners to be chamfered to ensure a good print quality.
    \item Print orientation - printed with the exposed lattice face facing upwards so that the mirror surface is self-supporting.
    \item Powder removal - open back design so that excess powder can be removed easily.
    \item Post-processing - to consider design for manufacture requirements, CNC machining and SPDT, as well as optional thermal treatments, such as HIP, to reduce porosity.
\end{enumerate}

\section{Design Process}
\label{sec:DesignProcess}

\begin{figure}[htbp]
\begin{center}
\includegraphics[width = 0.75\textwidth]{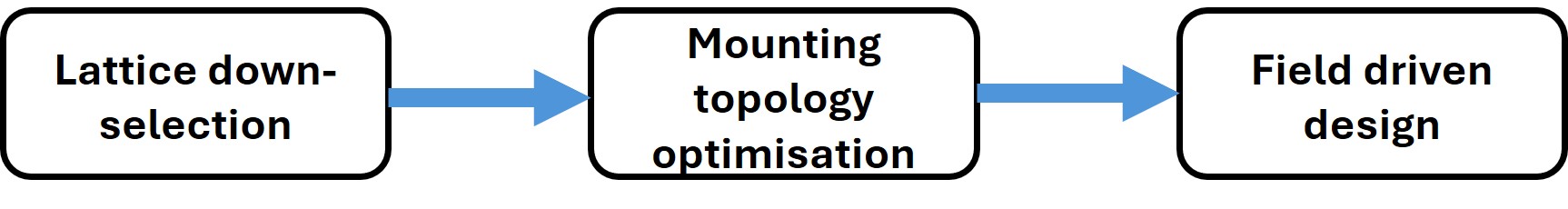}
\end{center}
\caption{Design process steps.} 
\label{fig:designprocessdiagram}
\end{figure}

The design process involved three steps (Figure~\ref{fig:designprocessdiagram}), first, a lattice down-selection for the internal mirror volume. Second, with the lattice down-selection complete, a mounting structure was developed to transfer loads away from the mirror surface. Finally, the mirror body was combined with the mounting structure for a final optimisation using a field-driven design approach~\cite{Morris2024}.

\subsection{Lattice design}
\begin{figure}[htbp]
\begin{center}
\includegraphics[width = 0.95\textwidth]{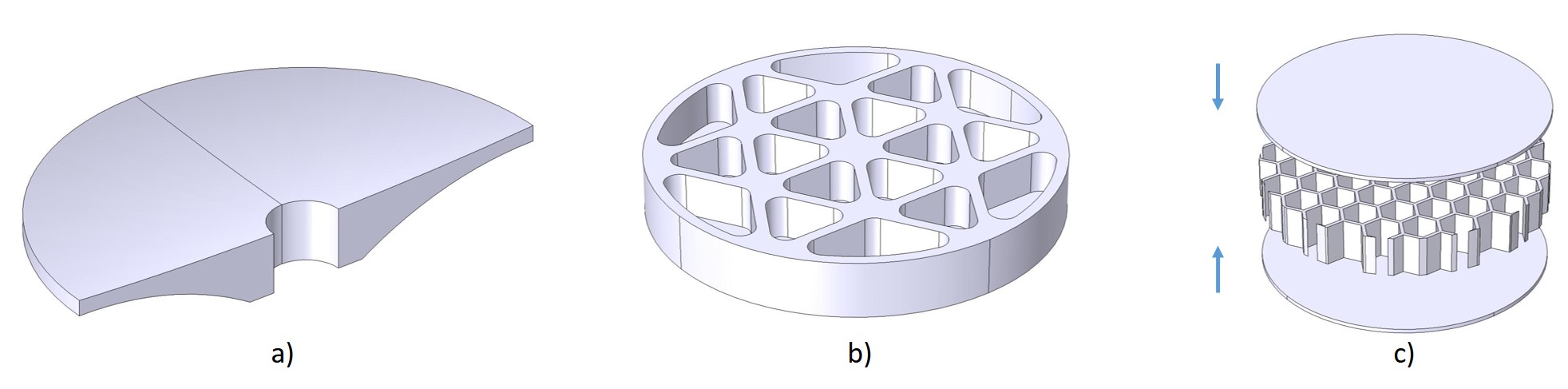} 
\end{center}
\caption{Conventional mirror design light-weighting methods: a) contoured back, b) open back and c) sandwich. Image credit: \textit{Atkins et al. (2021)} \cite{AMCookBook}.}
\label{fig:conventionalmirrorexample}
\end{figure}

Lattice-based AM mirrors have demonstrated superior strength-to-weight ratios compared to their conventional design counterparts \cite{zhang2022review}, (Figure~\ref{fig:conventionalmirrorexample}). To accelerate the design process, a reduced set of lattice types were compared against each other and optimised using FEA to assess their deformation under the loading conditions of SPDT. The lattices chosen for this comparison were selected from the paper by \textit{Lister et al. (2024)} \cite{lister2024design} which was particularly relevant due to the similarity in mirror geometry and size. 

\begin{figure}[htbp]
\begin{center}
\includegraphics[width = 0.6\textwidth]{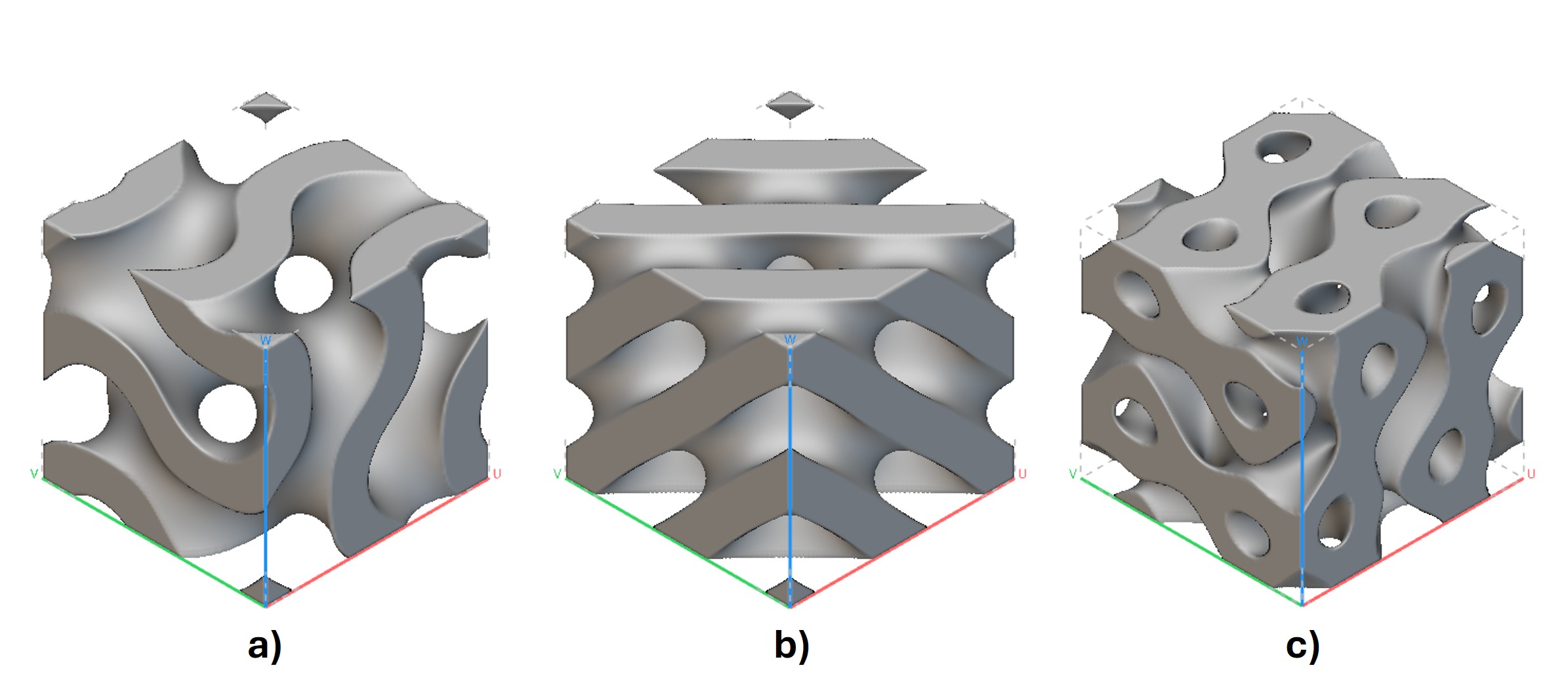} 
\end{center}
\caption{TPMS unit cell CAD: a) gyroid, b) diamond, c) split-P.}
\label{fig:TPMS Unit Cell Examples}
\end{figure}

\begin{figure}[htbp]
\begin{center}
\includegraphics[width = 0.475\textwidth]{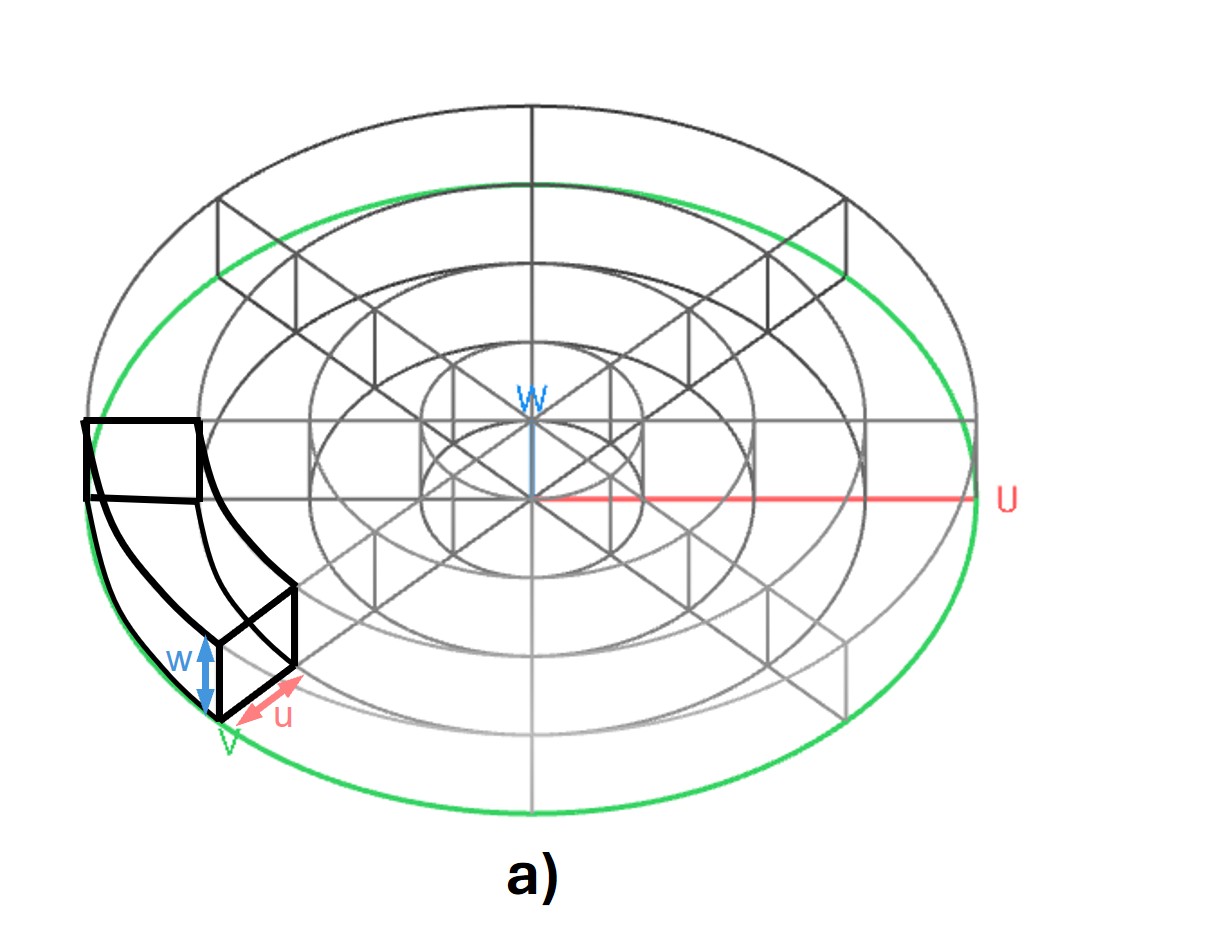}
\includegraphics[width = 0.475\textwidth]{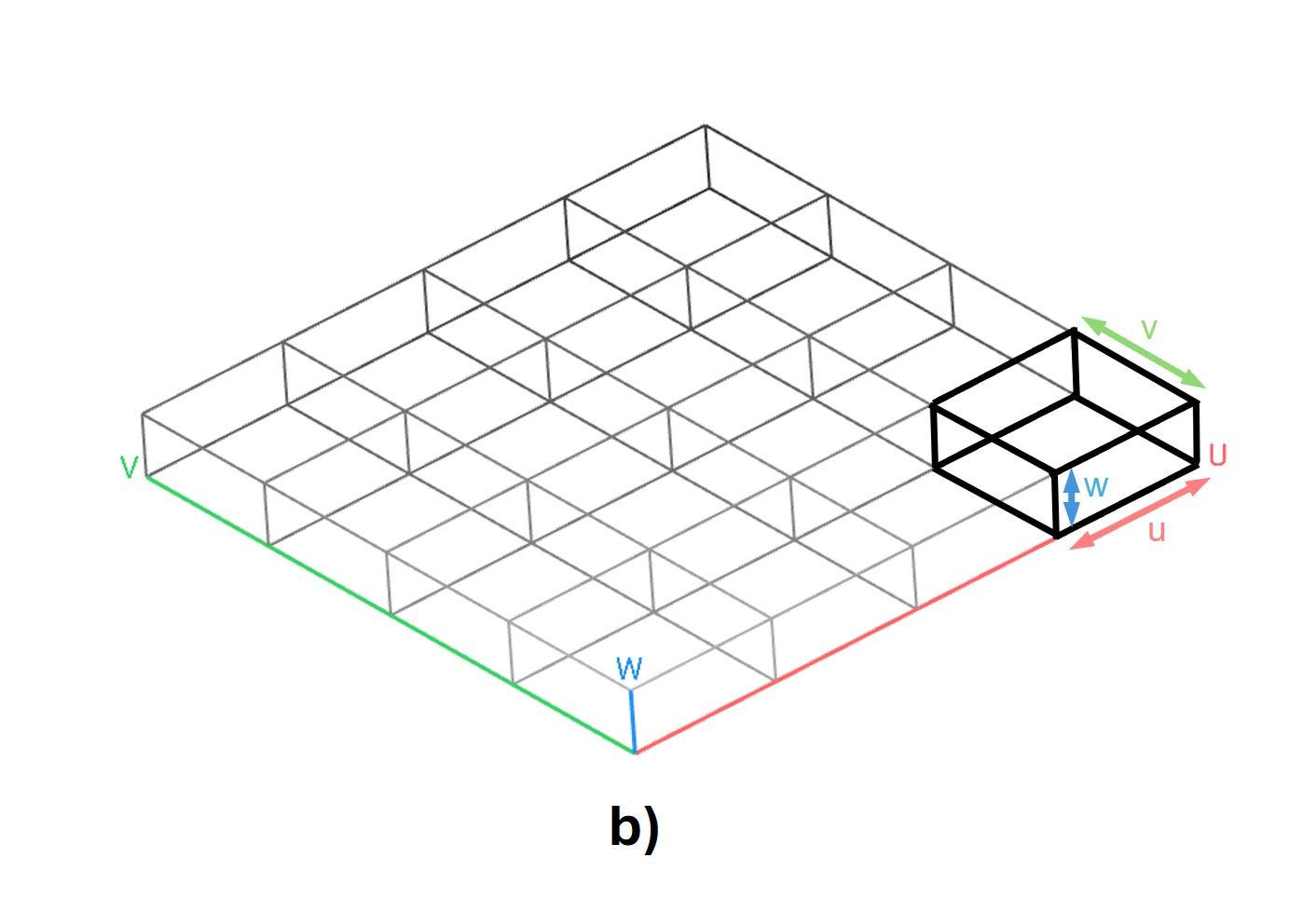}
\end{center}
\caption{Example cell maps: \textit{left} a cylindrical cell map, with parameters, unit cell radius (u), arc count (v) and unit cell height (w); \textit{right} a cubic cell map, with parameters, unit cell width (u); unit cell depth (v) and unit cell height (w).}
\label{fig:CellMapExamples}
\end{figure}

An implicit modelling CAD software was used to create the lattice structures tested, which meant that the geometry was defined by mathematical equations rather than each boundary being defined explicitly. This allowed for the direct creation of intricate geometries but also faster and simpler design modification.Within the TPMS family of lattices, the gyroid, diamond, and split-P types were highlighted by \textit{Lister et al. (2024)} \cite{lister2024design} as the best candidates to minimise print-through; the lattice unit cells are shown in Figure~\ref{fig:TPMS Unit Cell Examples}. A unit cell can be mapped to a volume in three coordinate frames: cubic, cylindrical, and spherical. The spherical cell map was excluded from this study as it did not match the flat profile of the mirror design; as such, only the cylindrical and cube cell maps were considered, and Figure~\ref{fig:CellMapExamples} defines the unit cell parameters for each map. Cylindrical cell maps are characterized by the unit cell's radius (size in the radial direction), height (size in the vertical direction), and the number of unit cells around the circumference. In cubic cell maps, unit cells are defined by their width (horizontal size), height (vertical size), and depth. These parameters are visualised in Figure~\ref{fig:CellMapExamples}. The final parameter to be optimised was lattice wall thickness which, as discussed in Section~\ref{sec:DesignSpec}, was limited to values of $\ge$ \SI{1}{\milli\meter}.

\subsubsection{FEA setup}

 Mesh parameters suggested by literature were used. Triangular meshes are more flexible than rectangular meshes and, therefore, better suited to the complex lattice profiles tested, resulting in more accurate simulations. An edge length of double the smallest feature size was recommended by \textit{Xinyin et al. (2021)} \cite{jia2021research}, additionally, a tolerance on the edge length of 10\% of the smallest feature size. With the smallest feature size being \SI{1}{\milli\meter} to guarantee a successful build, the mesh edge length was set to \SI{2}{\milli\meter} with a tolerance of \SI{0,2}{\milli\meter}.

\begin{figure}[htbp]
\begin{center}
\includegraphics[width = 0.95\textwidth]{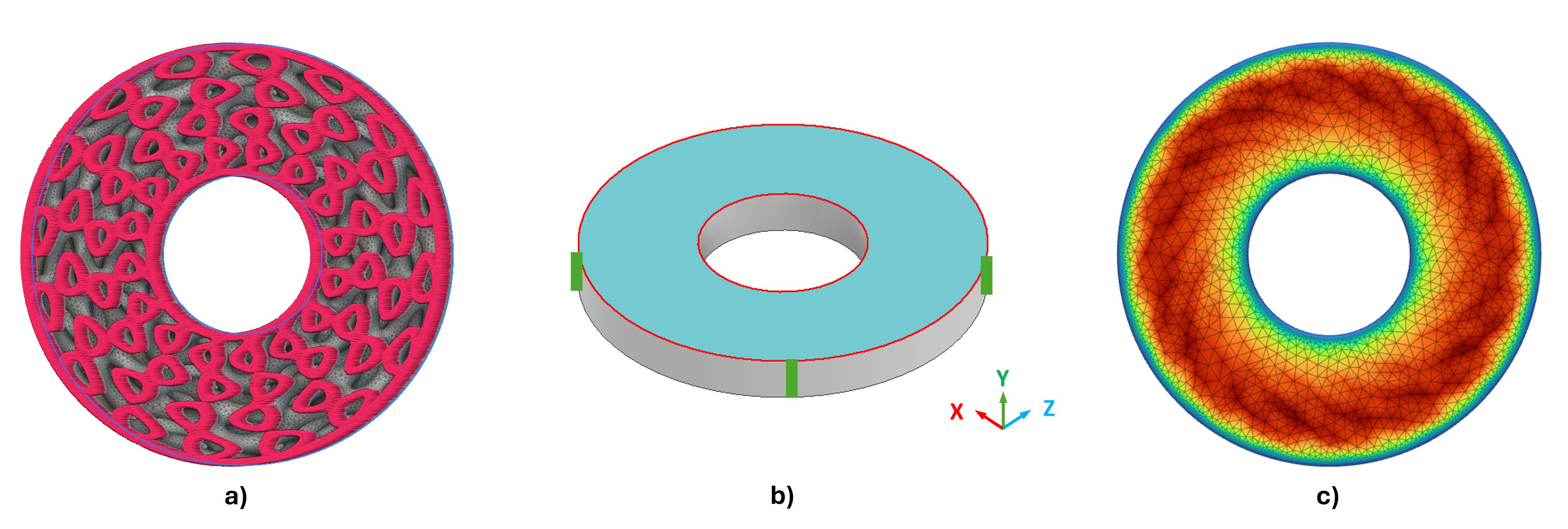}
\end{center}
\caption{FEA loading and boundary conditions: a) uniform surface pressure surface shown in blue; x and z axes restrictions shown in green. b) base restriction applied to a split-P cylindrical mapped lattice structure; c) Diamond TPMS displacement overlay plot with red and orange indicating displacement into the page and blue indicating zero displacement.}
\label{fig:FEASetUp}
\end{figure}

To replicate the loads experienced by the AM mirror during SPDT the following conditions were used for the FEA simulations. Firstly, the base was fixed in the y-axis as shown in Figure~\ref{fig:FEASetUp} a). Four vertical edges \SI{90}{\degree} apart were fixed in the x and z axes depending on which edges they were aligned with to constrain the model against twisting, as shown in Figure~\ref{fig:FEASetUp}. The rear face was constrained in the y-axis as this face would be in contact with the fixture plate. A uniform pressure of \SI{3500}{\pascal} was applied to the mirror surface; this value is a heritage polishing pressure used in previous design studies. In this study, the relative displacements are important, not the absolute values. The results of these boundary and loading conditions are visualised as a displacement overlay plot for a diamond TPMS lattice in Figure~\ref{fig:FEASetUp} c).

\subsubsection{Cylindrical cell map}
To accelerate the design phase, preliminary simulations evaluating cell map parameters and computational time were run to identify optimal baseline values for each parameter, balancing relatively low computational times against mass reduction values close to 60\%. These preliminary simulations began with cylindrical cell map lattices, initially varying the lattice wall thickness, followed by cell radius, cell height, and finally arc count. Testing of the rectangular cell map followed with lattice wall thickness tests, then cell depth, cell width, and finally cell height. For cylindrical cell map configurations, a cell radius and cell height of \SI{12}{\milli\meter}, arc count of 8, as well as a lattice wall thickness of \SI{1,4}{\milli\meter} were determined to be the most suitable baseline due to the relatively low computational time whilst achieving a mass reduction close to the targetted 60\% at 64\%, 67\% and 62\% for the gyroid, diamond and split-P lattices respectively. These preliminary tests also showed that setting the unit cell radius and unit cell height to the same value resulted in lower average surface deformation, therefore, this condition was kept for all subsequent tests. This set of values then determined the range of values tested for each lattice type. The upper limit of cell radius and cell height values tested were based on the values at which 70\% mass reduction was exceeded. Whereas, the upper limit of arc count and lattice wall thickness values tested corresponded to the value at which the mass reduction dropped below the targeted 60\% threshold. The lower limits were then set using the inverse of these, keeping the achieved mass reduction value between 60\% and 70\%, except for the lower limit for lattice wall thickness, which was set to the \SI{1}{\milli\meter} limit. Table~\ref{tab:CylindricalCellMapTestedParameters} shows the range of parameter values tested for the three lattice types.

\begin{table}[htbp]
\caption{Range of values tested for the cylindrical cell map gyroid, diamond and split-P parameters.} 
\label{tab:CylindricalCellMapTestedParameters}
\begin{center}       
\begin{tabular}{|l|l|l|l|l|}
\hline
\textbf{Lattice type} & \textbf{\begin{tabular}[c]{@{}l@{}}Lattice wall  \\ thickness {[}mm{]}\end{tabular}} & \textbf{\begin{tabular}[c]{@{}l@{}}Unit cell \\ radius {[}mm{]}\end{tabular}} & \textbf{\begin{tabular}[c]{@{}l@{}}Unit cell \\ height {[}mm{]}\end{tabular}} & \textbf{Arc count} \\ \hline
\textbf{Gyroid} & 1 - 2 & 8 - 16 & 8 - 16 & 6 - 12 \\ \hline
\textbf{Diamond} & 1 - 2 & 8 - 16 & 8 - 16 & 6 - 12 \\ \hline
\textbf{Split-P} & 1 - 2 & 8 - 16 & 8 - 16 & 6 - 12 \\ \hline
\end{tabular}
\end{center}
\end{table} 

\begin{figure}[htbp]
\begin{center}
\includegraphics[width=0.85\linewidth]{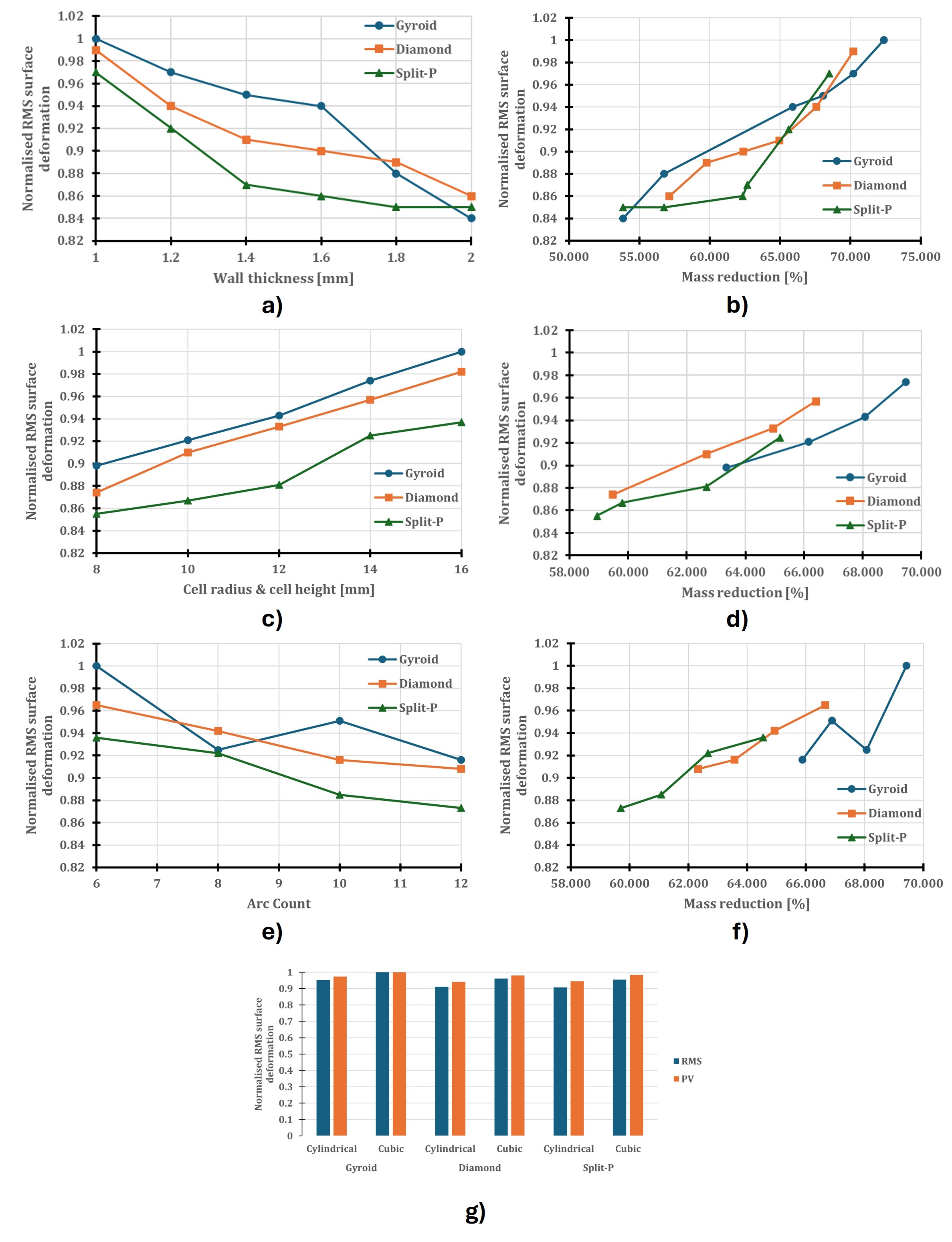}
\end{center}
\caption{Normalised cylindrical cell map configuration lattice down-selection plots: a) the effect of increasing lattice wall thickness on the RMS y-axis surface deformation; b) surface deformation represented as a mass reduction percentage by varying wall thickness; c) the effect of increasing unit cell radius and height on the RMS y-axis surface deformation; d) surface deformation represented as a mass reduction percentage by varying unit cell radius and height; e) the effect of increasing arc count on the RMS y-axis surface deformation; f)  surface deformation represented as a mass reduction percentage by varying arc count; g) Cylindrical and cubic cell map baseline lattice structures comparison of RMS and PV (peak to valley) surface deformation.}
\label{fig:CombinedLatticeDownselectionPlots}
\end{figure}

The first series of simulations trialled each parameter at an interval of 2 units, which reduced the total computational time, whilst highlighting the sub-range of values to focus on for each parameter. Two simulations were run per parameter, with an average taken to improve accuracy. Subsequent tests would identify the optimal value within these sub-ranges. The example FEA deformation in Figure~\ref{fig:FEASetUp} c) highlights the region of maximum deformation on the AM mirror surface matching the profile of the gaps in the diamond TPMS lattice structure beneath. The magnitude of deformation decreases towards the walls until it becomes near zero due to the additional support. Based on the first results delivering the anticipated deformation geometry, the FEA configuration was deemed suitable to evaluate all parameters.

From Figures~\ref{fig:CombinedLatticeDownselectionPlots} a) and b), it is evident that the split-P lattices had the lowest RMS surface deformation for all lattice wall thicknesses tested except for \SI{2}{\milli\meter} lattice wall thickness tests. This trend continued with the cell radius and height as well as the arc count simulation data as visualised in Figures~\ref{fig:CombinedLatticeDownselectionPlots} c) to f) with the split-P lattices resulting in less surface deformation than the diamond and gyroid lattices for the tested values. Additionally, the split-p lattice had a greater mass reduction when comparing equivalent lattice parameters. 

\subsubsection{Cubic cell map}

The cubic cell map adopted the same testing approach as the cylindrical with a cell width and cell depth of \SI{12}{\milli\meter}, cell height of 8, as well as a lattice wall thickness of \SI{1,4}{\milli\meter} chosen as the baseline parameters. The combination of tested parameters are shown in Table~\ref{tab:cubicCellMapTestedParameters}. However, it was found that the cubic cell map lattices performed worse than their cylindrical cell map counterparts, as shown by comparing the data points in Figure~\ref{fig:CombinedLatticeDownselectionPlots} g). Therefore it was decided to cease simulating the cubic lattices and to focus on additional simulations on the cylindrical split-P and diamond-based lattices.

\begin{table}[htbp]
\caption{Range of values tested for the rectangular cell map gyroid, diamond and split-P parameters.} 
\label{tab:cubicCellMapTestedParameters}
\begin{center}       
\begin{tabular}{|l|l|l|l|l|}
\hline
\textbf{Lattice type} & \textbf{\begin{tabular}[c]{@{}l@{}}Lattice wall  \\ thickness {[}mm{]}\end{tabular}} & \textbf{\begin{tabular}[c]{@{}l@{}}Unit cell \\ width {[}mm{]}\end{tabular}} & \textbf{\begin{tabular}[c]{@{}l@{}}Unit cell \\ depth {[}mm{]}\end{tabular}} & \textbf{\begin{tabular}[c]{@{}l@{}}Unit cell \\ height {[}mm{]}\end{tabular}} \\ \hline
\textbf{Gyroid} & 1 - 2 & 8 - 16 & 8 - 16 & 8 - 14 \\ \hline
\textbf{Diamond} & 1 - 2 & 8 - 16 & 8 - 16 & 8 - 14 \\ \hline
\textbf{Split-P} & 1 - 2 & 8 - 16 & 8 - 16 & 8 - 14 \\ \hline
\end{tabular}
\end{center}
\end{table}

\newpage

\subsubsection{Final lattice configuration}
\begin{figure}[t]
\begin{center}
\includegraphics[width = 0.55\textwidth]{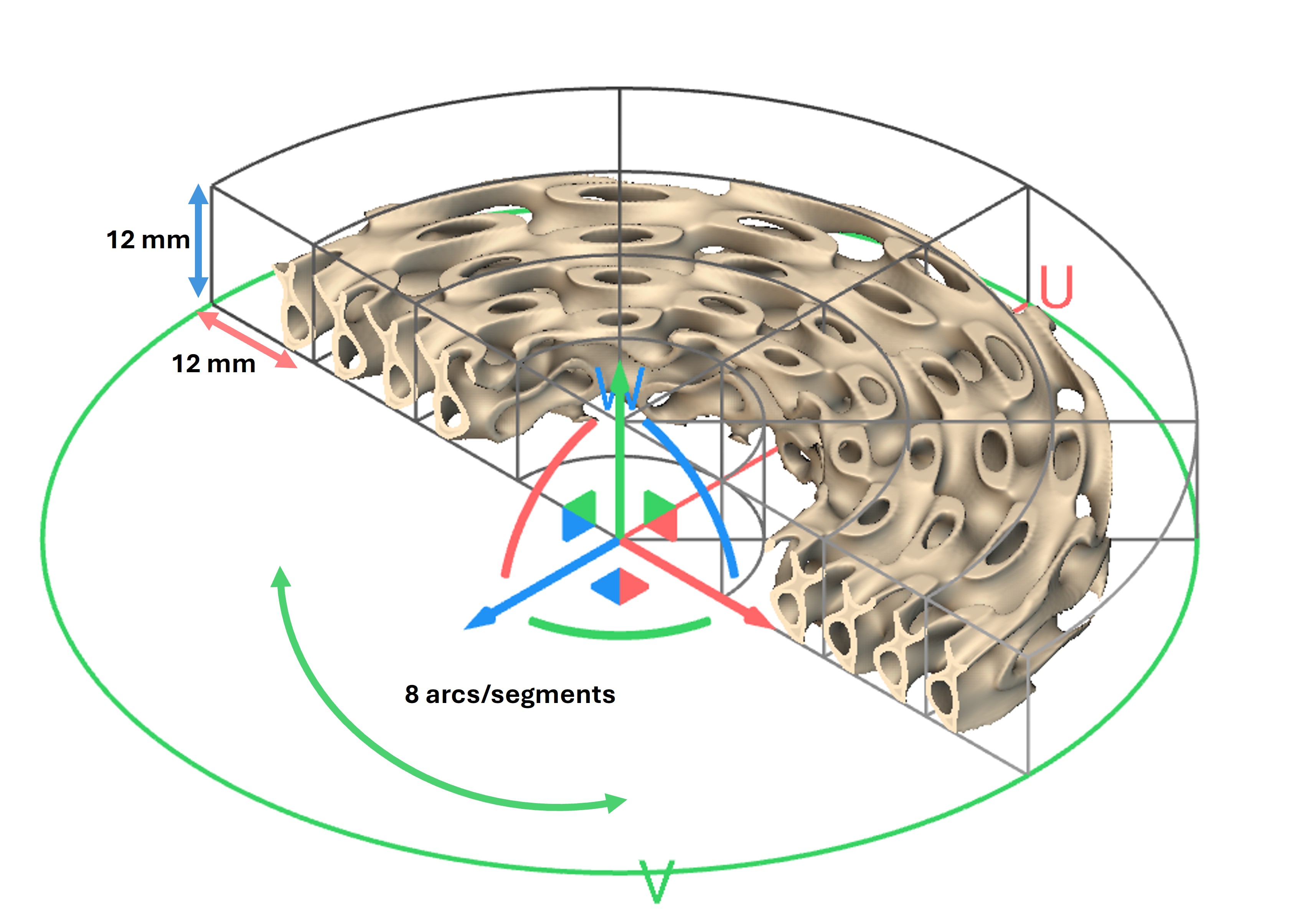}
\end{center}
\caption{Finalised lattice configuration shown within the finalised cell map.}
\label{fig:FinalLatticeConfiguration}
\end{figure}

The lattice configuration selected was the split-P cylindrical cell map with a cell radius and cell height of \SI{12}{\milli\meter}, an arc count of \si{8}, and a lattice wall thickness of \SI{1,4}{\milli\meter}, presented in Figure~\ref{fig:FinalLatticeConfiguration}. This achieved a mass of 64.2\% compared to a solid body equivalent. 

\subsection{Mounting structure design process}

\begin{figure}[htbp]
\begin{center}
\includegraphics[width = 0.95\textwidth]{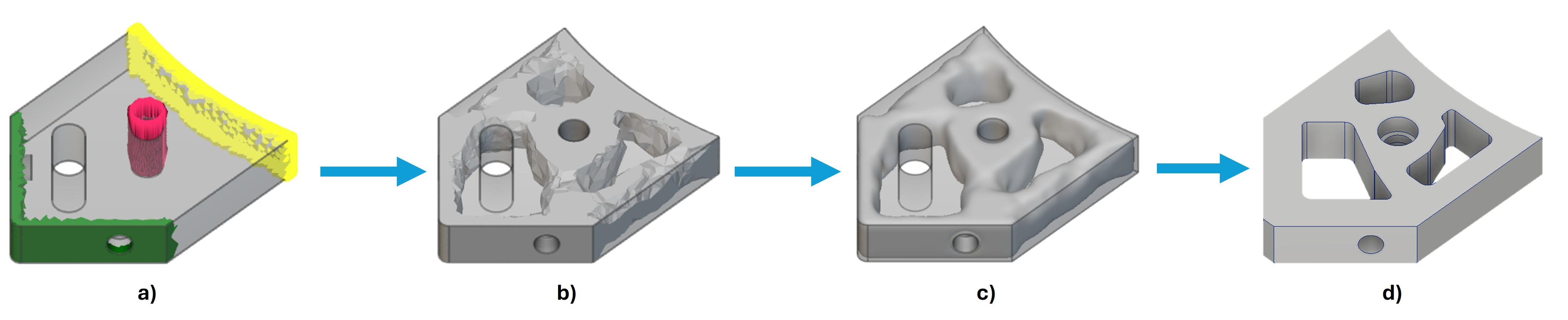}
\end{center}
\caption{ Mounting structure design flow: a) topology optimisation set up with the loaded face in yellow, the restrained fixture point in red, and the contained faces in green; b) raw topology optimisation output; c) smoothened topology optimisation output; d) simplified topology optimised mounting structure by overlay of the topology optimisation output in CAD.}
\label{fig:mountingtopop}
\end{figure}

The asymmetric layout of the PCB mounting rods within the CubeSat chassis posed an issue for the design of the mounting structure. If the clearance holes were to match the diameter of the rods, their location at each corner would need to be at three separate positions relative to the midpoint of the mirror. Requiring three separate designs and consequently three separate topology optimisation processes. To simplify, a single mounting design was developed that accommodated all three variations in the position of the PCB rod, which included a slot that allowed the four rods to pass through, as shown in Figure~\ref{fig:mountingtopop}.

To simulate the loading conditions during diamond turning a point moment was applied to the mirror mount interface. The moment arm extended to the centre of rotation of SPDT, and the force applied was an arbitrary albeit physical value, similar to the lattice down-selection; the individual displacement magnitudes were not relevant, rather a comparison between the different design iterations. As the simulations were optimising for minimal surface displacement during SPDT, the SPDT fixture points served as the restraints. This setup could then be used to run a 2D topology optimisation with the two requirements: that no material be added to the PCB rod clearance slot; and the outer edges be maintained to be butted against during the SPDT process. The results of this topology optimisation were overlayed onto the original CAD part, approximations of the cutout areas were then extruded the entire \SI{5}{\milli\meter} thickness of the part as shown in Figure~\ref{fig:mountingtopop}. Next, this mounting structure was integrated with the mirror walls and surface in CAD to be united with the optimised lattice structure, which, when complete, achieved a total mass reduction of 56\%.

\subsection{Field driven lattice design}
\begin{figure}[htbp]
\begin{center}
\includegraphics[width = 0.95\textwidth]{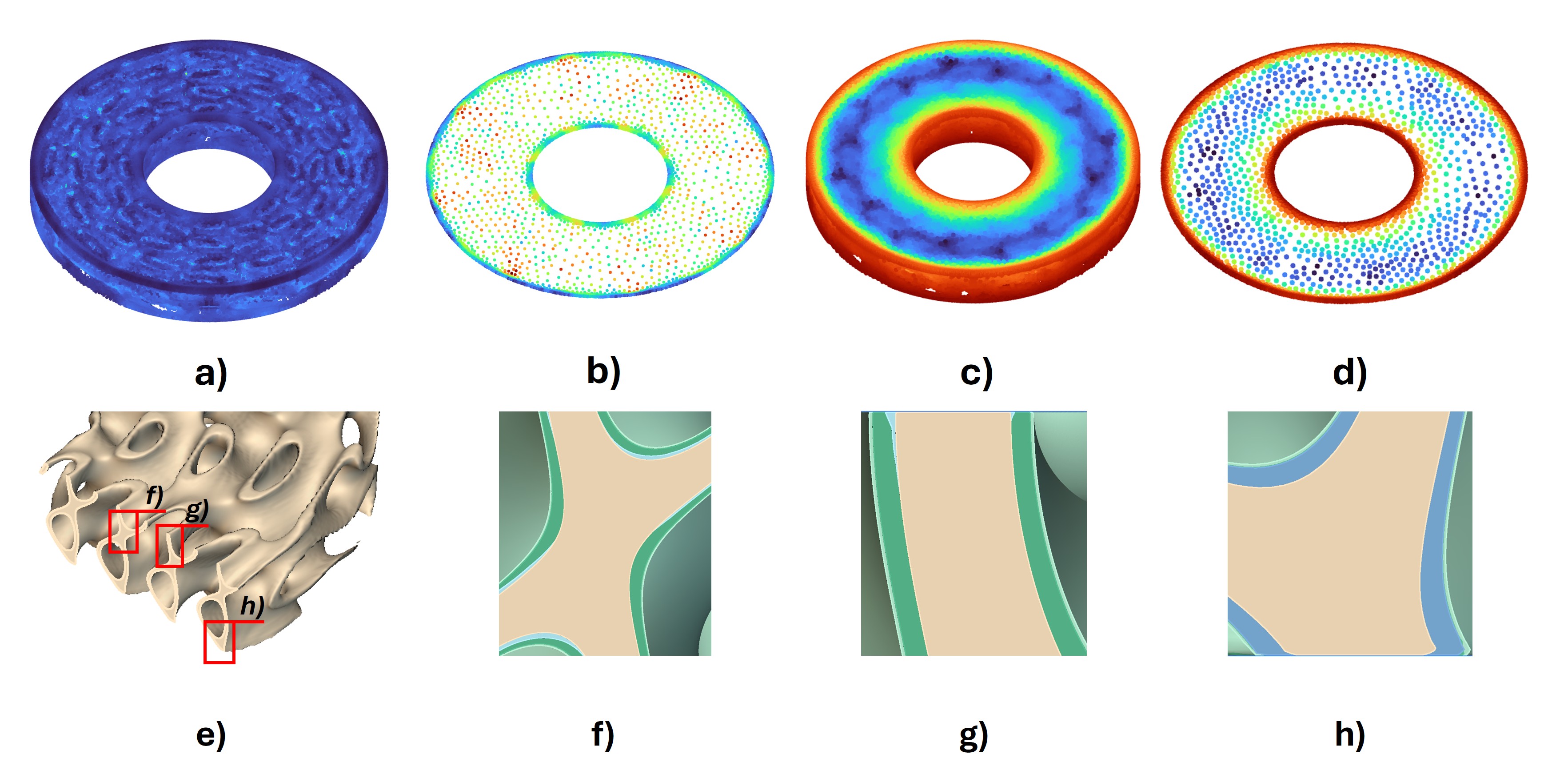}
\end{center}
\caption{ The four-point map variations used as inputs for the field-driven design lattice structures and the resultant field driven latices with increase wall thicknesses overlayed in section view: a) volume stress point map; b)  surface stress point map; c) volume displacement point map; d)  surface displacement point map; e) section view of the uniform thickness lattice highlighting the regions shown in f) to h); f) to h) uniform lattice wall thickness ramp lattice shown in tan; volume stress ramp lattice shown in light blue; optical surface stress ramp lattice shown in dark blue; volume displacement ramp lattice shown in light green; optical surface displacement ramp lattice shown in light dark green.}
\label{fig:fielddrivenpointmaps}   
\end{figure}

Once the mounting structure was finalised, a field-driven design approach was used to optimise the lattice thickness for uniform surface displacement using more representative geometry and boundary constraints. For example, the stress transferred across the mounting/mirror interface via the inclusion of the finalised mounting structure. A field-driven design inputs an analysis output back into a design process to further optimise for a given design objective, in this case, the y-axis surface deformation. The chosen parameter to feed into this optimisation was the lattice wall thickness, resulting in a ramp lattice which varied in thickness within a set range of values. Firstly, a set of boundary conditions was applied with a uniform pressure applied to the mirror surface, whilst the mirror base was restrained in the vertical (y) axis only, the mount corner faces restrained in their respective planes, and the SPDT fixture holes restrained in all axes. A FEA of the assembly was calculated using these boundary conditions, and a displacement point map of the resultant data was then generated, and the y-axis data points isolated from both the entire mirror body and the optical surface deformation. The same steps were followed using the Von Mises stress output of the FEA for an alternative ramp lattice option as a comparison. Figure~\ref{fig:fielddrivenpointmaps} shows the resultant scalar point maps, which were used as inputs for the lattice wall thickness variable, effectively increasing the thickness in areas of high y-axis deformation or stress. This thickness was limited to a range of \SI{1}{\milli\meter} to \SI{2}{\milli\meter} to ensure the part mass reduction was close to the targetted 60\% mass reduction.

\begin{figure}[htbp]
\begin{center}
\includegraphics[width=0.975\linewidth]{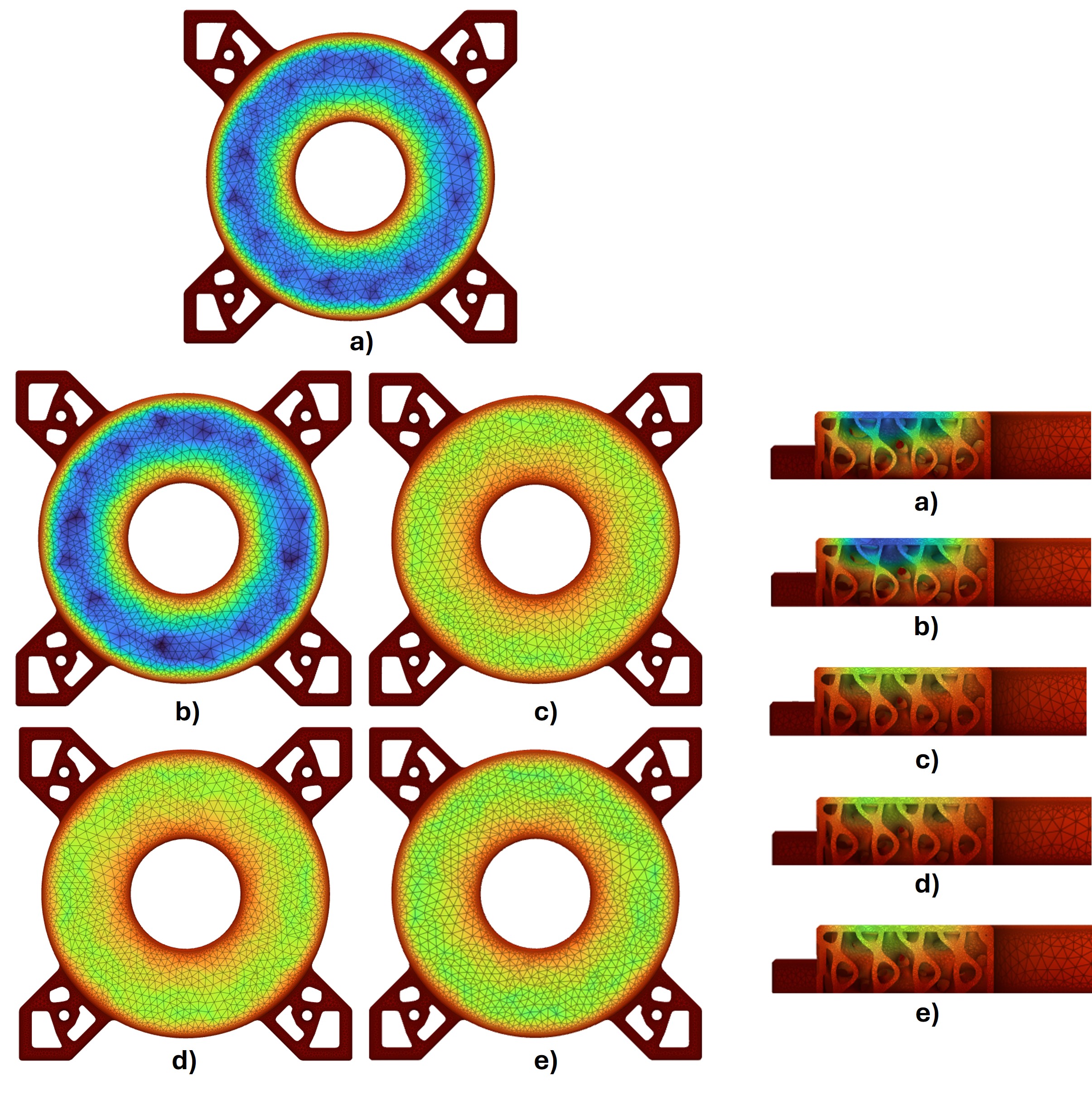}
\end{center}
\caption{y-axis displacements overlay plot of field driven lattices optical surface as well as section views: a) uniform lattice wall thickness; b) volume stress based ramp lattice; c) optical surface stress based ramp lattice; d) volume displacement based ramp lattice; d) optical surface displacement based ramp lattice. A standardised colour map is used for all shown overlay plots, with red indicating out-of-page displacement and green and blue indicating into-page displacement.}
\label{fig:Fielddrivenoverlayplots}
\end{figure}

Figures~\ref{fig:fielddrivenpointmaps} f) to h) show the resultant ramp lattices as well as a \SI{1,4}{\milli\meter} uniform thickness lattice overlayed to visualise the difference in lattice wall thicknesses. The volume stress-based ramp lattice is similar in profile to the uniform thickness lattice, and so minimal reduction in y-axis surface deformation was expected from this design iteration. The remaining ramps lattices showed more promise with greater variation relative to the uniform thickness lattice. This is confirmed in Figure~\ref{fig:Fielddrivenoverlayplots}, where the effectiveness of the ramp lattices is visualised in the y-axis displacement overlay plots and quantified in Table~\ref{tab:FieldDrivenSurfaceDeformation}. The optical surface stress-based ramp lattice performed the best with the least RMS surface deformation in the y-axis, whilst achieving a mass reduction value close to the target 60\%.

\begin{table}[htbp]
\caption{Field-driven lattices surface deformation results, selected field ramp highlighted.} 
\label{tab:FieldDrivenSurfaceDeformation}
\begin{center}       
\begin{tabular}{|p{3cm}|p{2.5cm}|p{2.5cm}|p{3cm}|p{3cm}|} 
\hline
\rule[-1ex]{0pt}{3.5ex} \multirow{2}{*}{\textbf{Field ramp}} & \multicolumn{2}{|c|}{\textbf{Surface displacement - y axis*}} & \multicolumn{2}{|c|}{\textbf{Mass reduction \%}} \\
\cline{2-5}
\rule[-1ex]{0pt}{3.5ex}   & \textbf{RMS} & \textbf{Max.} & \textbf{Mirror body} & \textbf{Mirror + mounts} \\
\hline
\rule[-1ex]{0pt}{3.5ex} \textbf{Uniform} & \SI{1,00}{} & \SI{1,00}{} & \SI{62,67}{} & \SI{57.08}{} \\
\hline
\rule[-1ex]{0pt}{3.5ex} \textbf{Volume - stress} & \SI{0,98}{} & \SI{1,05}{} & \SI{60,95}{} & \SI{55,73}{}\\
\hline
\rule[-1ex]{0pt}{3.5ex}  \cellcolor[HTML]{CCFFCC}\textbf{Surface - stress} & \cellcolor[HTML]{CCFFCC}\SI{0,64}{} & \cellcolor[HTML]{CCFFCC}\SI{0,58}{} & \cellcolor[HTML]{CCFFCC}\SI{55,54}{} & \cellcolor[HTML]{CCFFCC}\SI{51,11}{}\\
\hline
\rule[-1ex]{0pt}{3.5ex}   \textbf{Volume - disp.} & \SI{0,70}{} & \SI{0,57}{} & \SI{50,47}{} & \SI{50,21}{}   \\
\hline
\rule[-1ex]{0pt}{3.5ex}   \textbf{Surface - disp.} & \SI{0,67}{} & \SI{0,60}{} & \SI{55,58}{} & \SI{51,14}{}  \\
\hline
\multicolumn{5}{l}{}\\
\multicolumn{5}{l}{* - normalised data to to uniform field ramp.}\\
\end{tabular}
\end{center}
\end{table} 

\newpage
\section{Manufacture}
\label{sec:Manufacture}
\begin{figure}[htbp]
\begin{center}
\includegraphics[width=1\linewidth]{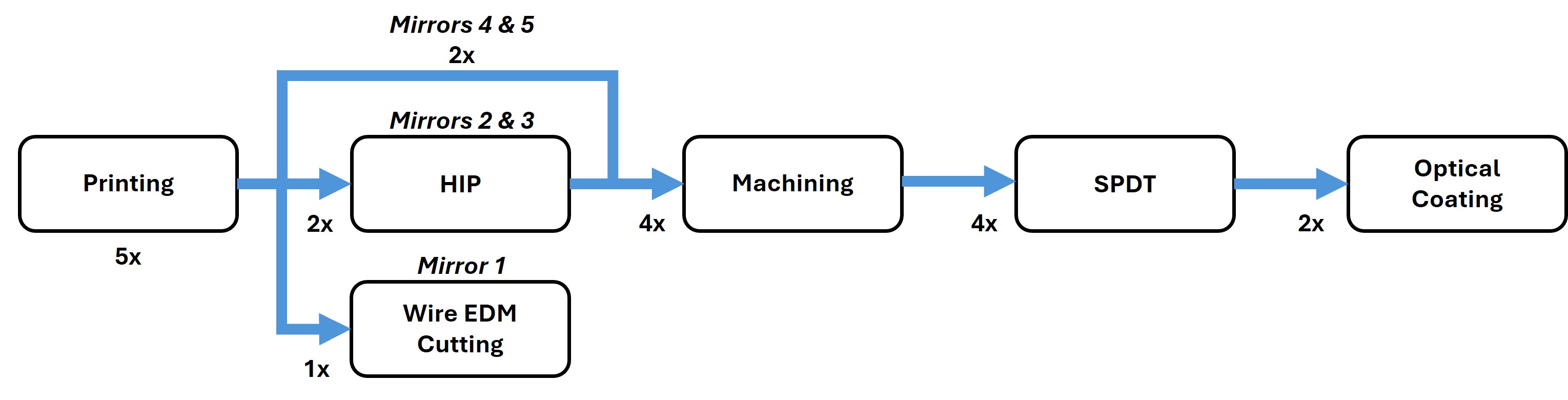}
\end{center}
\caption{Manufacturing process steps with the number of mirrors subject to each process shown.}
\label{fig:ManufacturingProcessSteps}
\end{figure}

Manufacturing began with the printing of five mirrors. Each mirror had different combinations of post-processing steps applied to compare the resultant surface quality. Two of the five mirrors were subjected to HIP; the same two mirrors, along with two others, were then subjected to machining. SPDT was performed to reduce surface roughness and form error in preparation for the application of an optical coating. An optical coating was applied to one of the HIP-processed mirrors and one without. The final mirror was sectioned using wire electrical discharge machining (wire EDM), for XCT analysis. These steps are summarised in Figure~\ref{fig:ManufacturingProcessSteps}.

\subsection{Fabrication}   

\begin{figure}[htbp]
\begin{center}
\includegraphics[width=0.95\linewidth]{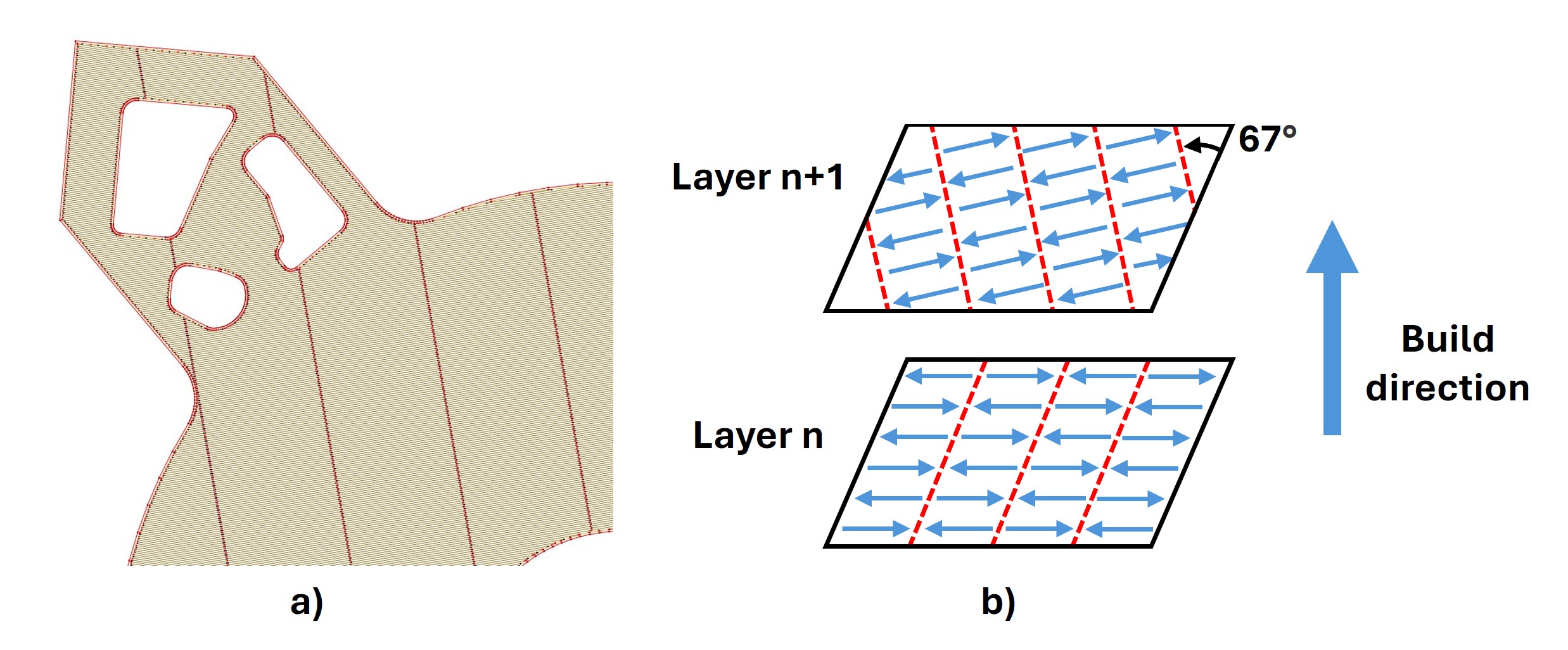}
\end{center}
\caption{Laser scan strategy used during the printing process. Image credits: a) M. Sanchez and b) adapted from \textit{Lu et al. (2022)} \cite{lu2022modelling}.}
\label{fig:printscanstrategy}
\end{figure}  

\begin{figure}[htbp]
\begin{center}
\includegraphics[width=0.95\linewidth]{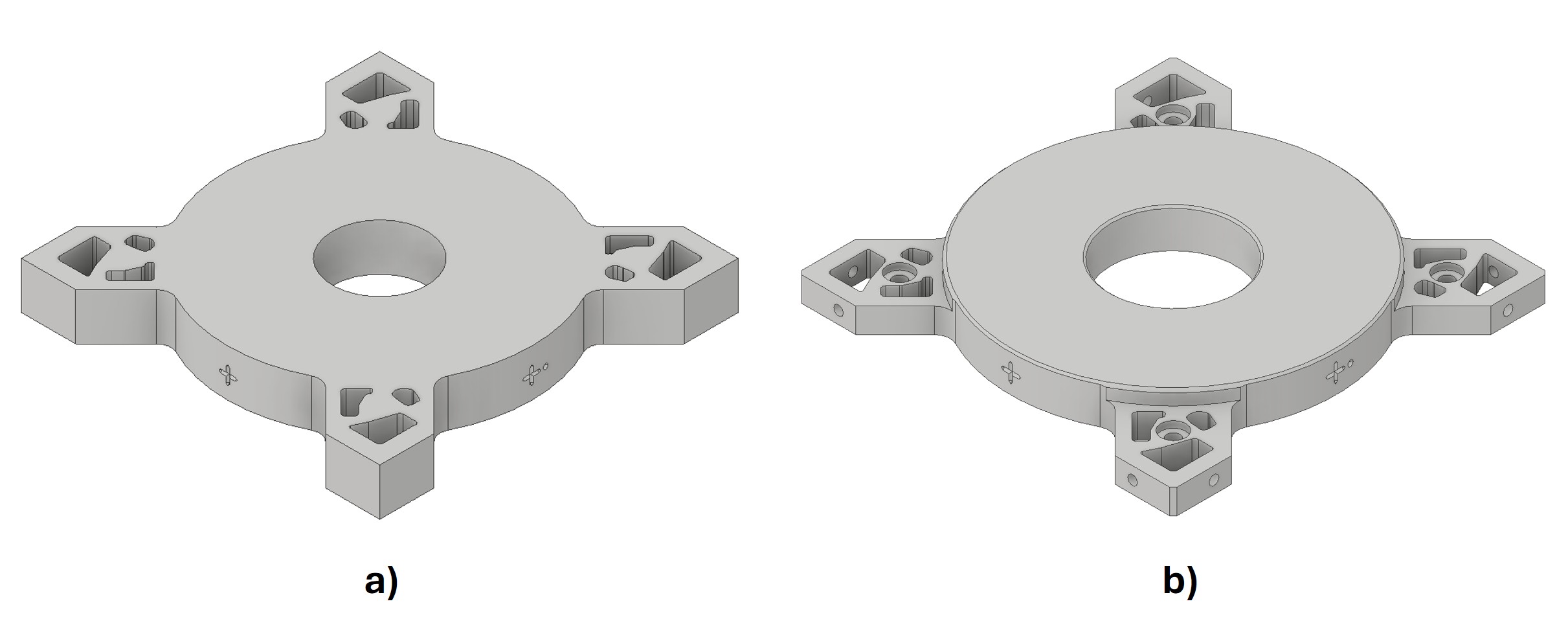}
\end{center}
\caption{Differences between the as-printed (\textit{left}) and as-machined (\textit{right}) designs represented in CAD.}
\label{fig:asprintedvsasmachinedCAD}
\end{figure}

Five mirrors were printed, all in AlSi10Mg aluminium alloy, using the L-PBF printing process via an external AM Bureau's SLM 500L. The mirrors were printed with the exposed lattice facing upwards, and the laser scan strategy used is shown in Figure~\ref{fig:printscanstrategy}. Excess material was added to all surfaces to be machined, including the base. Furthermore, \SI{2}{\milli\meter} of material was added to the mirror surface  with the perimeter increased from \SI{97}{\milli\meter} square to \SI{98,5}{\milli\meter} square as shown in Figure~\ref{fig:asprintedvsasmachinedCAD}. The mounts had \SI{7,0}{\milli\meter} of material added to their upper surface to bring them level with the as-printed mirror surface. Additionally, the central bore was printed with a diameter of \SI{26}{\milli\meter} as it was used to centre the part during machining. After printing, two of the four mirrors were subject to HIP before progressing onto machining.  

\subsection{Machining}
\begin{figure}[htbp]
\begin{center}
\includegraphics[width=0.95\linewidth]{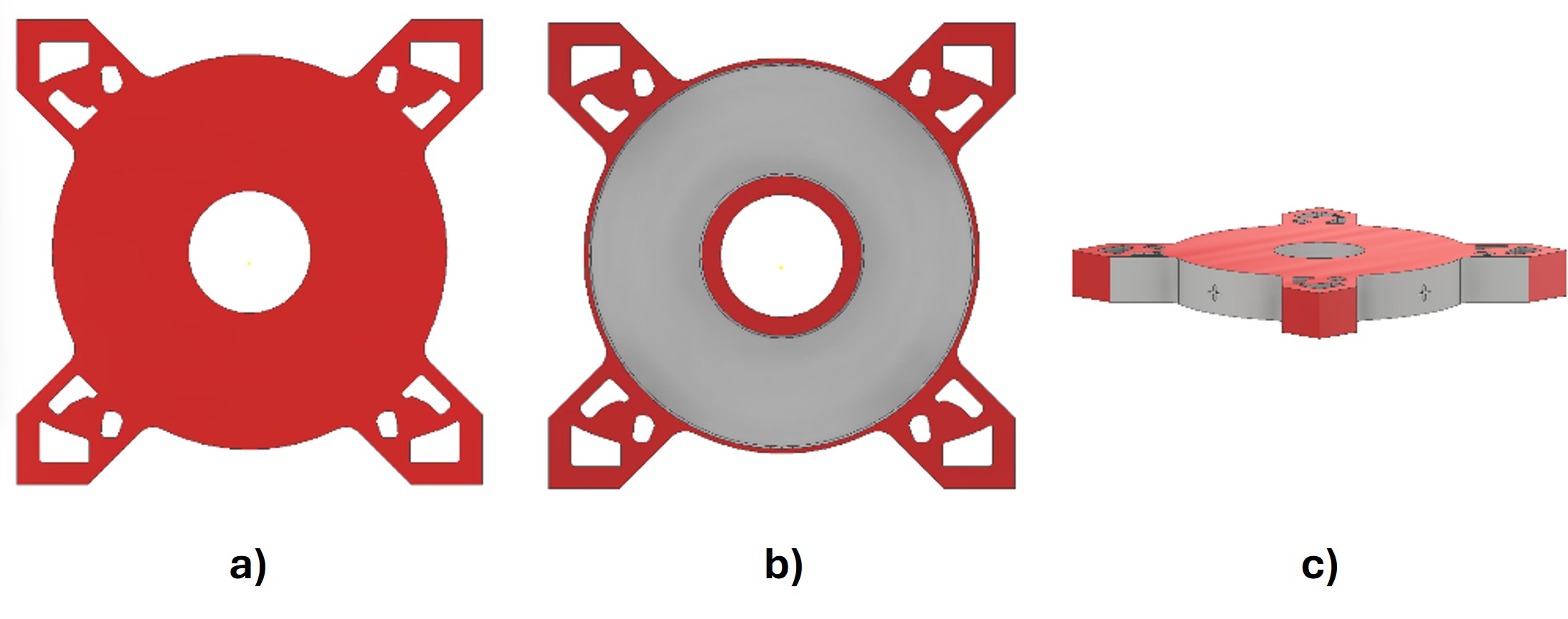}
\end{center}
\caption{CAD highlighting the machined surfaces on the as-printed part in red: a) top surface; b) lower surface with the lattice not shown as it was not machined; c) side profiles.}
\label{fig:MachiningSurfaces}
\end{figure}

\begin{figure}[htbp]
\begin{center}
\includegraphics[width=0.95\linewidth]{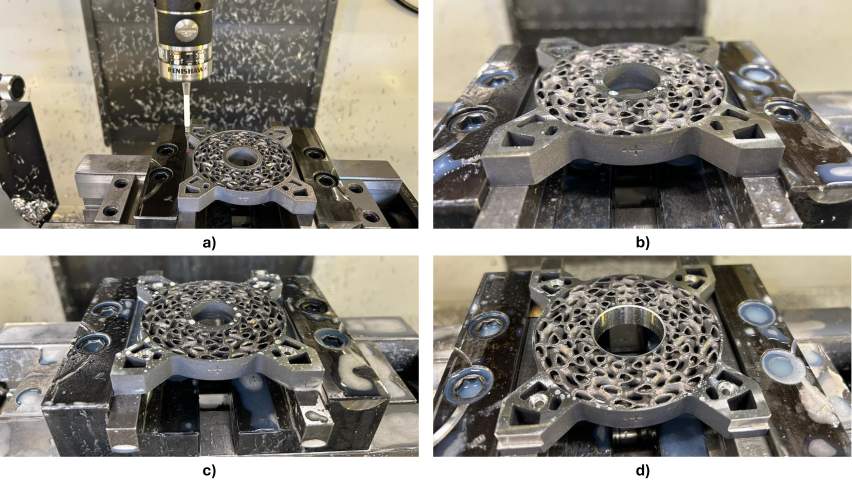}
\end{center}
\caption{Photos of the rough machining stages of the back-face: a) probing of the part to determine the profile and positioning in relation to the 5-axis milling machine; b) the back face after being machined flat; c) SPDT fixture holes drilled into the part; d) corner faces machined flat and chamfers added.}
\label{fig:machiningsteps}
\end{figure}

The printed parts were subjected to a series of in-house machining steps at UK ATC to achieve the dimensional tolerances required for SPDT. Figure~\ref{fig:MachiningSurfaces} shows the differences between the as-printed and as-machined parts. Firstly, a reference surface was required; this was assigned as the base of the part and consequently the first surface to be machined, with a flatness tolerance of \SI{0,1}{\milli\meter}. Next, the SPDT fixture holes, including the counterbores, were drilled into the back face. The last operation performed on the back-face was the machining of the corner faces, again to a flatness tolerance of \SI{0,1}{\milli\meter} and chamfers added to all sharp corners. These steps are shown in Figure~\ref{fig:machiningsteps}. The same steps were then applied to the top half of the part, with a final step of drilling and tapping the side-mounted fixture holes.   

\subsection{Single point diamond turning}
\begin{figure}[htbp]
\begin{center}
\includegraphics[width=0.95\linewidth]{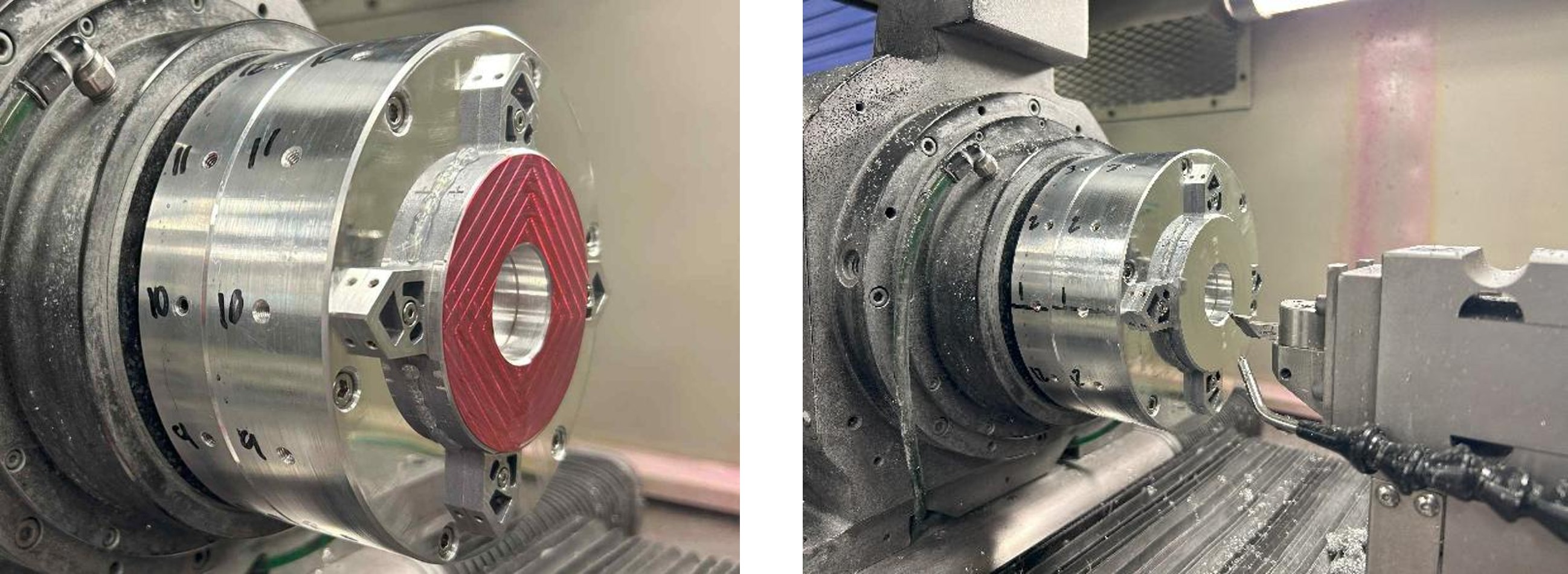}
\end{center}
\caption{Mirror mounted on a Nanotech 250UPL 3-axis ultra-precision lathe: left, Mirror \#4 with its optical surface coloured in to highlight a uniform cut; right, Mirror \#4 after the rough cuts had been performed.}
\label{fig:SPDTlathe}
\end{figure}

The four machined mirrors, two of which had undergone HIP and two without, were single-point diamond turned at Durham University using a Nanotech 250 UPL 3-axis ultra-precision lathe. Each mirror was first balanced for minimal vibration during the SPDT process, as the mirrors were rotated at 2000 revolutions per minute (rpm) during turning.  This was achieved by adjusting 12 screws located circumferentially around the edge of the fixture plate. A series of rough cuts were first made at \SI{10}{\micro\meter} depth, with up to ten of these cuts before fine cuts of \SI{0,5}{\micro\meter} were made, a mirror is shown mounted to the lathe in Figure~\ref{fig:SPDTlathe}.

\begin{figure}[htbp]
\begin{center}
\includegraphics[width=0.95\linewidth]{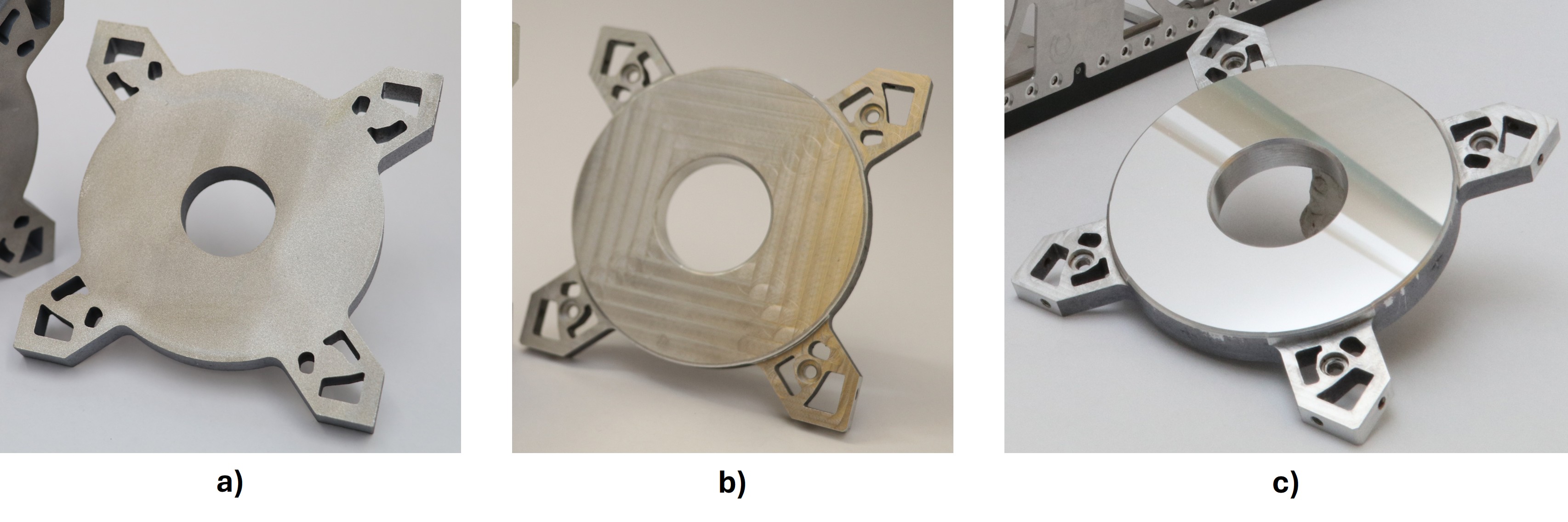}
\end{center}
\caption{Mirror prototype shown at the different stages of manufacture: a) as-printed, b) as-machined, c) post SPDT.}
\label{fig:ManufacturingStepsComparison}
\end{figure}

Post SPDT, qualitative differences between the HIP and non-HIP applied mirrors were observed. The non-HIP applied mirrors had greater reflectivity but also porosity in structured diamond patterns across the optical surface, whereas the HIP mirrors had reduced porosity but greater surface roughness. An internal evaluation was next performed using XCT to quantify internal porosity. Additionally, an external evaluation was performed using microscopy and interferometry to form a quantitative comparison of the differences between the HIP and non-HIP mirrors. A comparison of the mirrors throughout the manufacturing stages is shown in Figure~\ref{fig:ManufacturingStepsComparison}.

\newpage

\section{Evaluation - Internal, XCT}
\label{sec:Evaluation - internal, XCT}
XCT is a non-destructive imaging technique which has become a popular method for assessing the internal defects within AM parts\cite{XCTAMOverview}, specifically the location and volume of internal pores. The machine used for XCT analysis within this case study utilised a cone beam scanner. This is where the targeted part is rotated around a fixed axis between an X-ray source and a detector. A cone beam of X-rays emitted from the source passes through the part, with some of these X-rays absorbed depending on the variation in density through the part, revealing the internal structure. A projection of the part is then formed on the detector, with over a thousand images taken as the part is rotated, which are combined digitally to create a 3D representation of the X-ray attenuation. The 3D representation is a greyscale: the denser the material, the lighter the shade assigned to the pixel at that location.

\textit{Westik et al. (2023)} \cite{westsik2023design} discussed the challenges posed by an XCT interpretation by analysing the structure of a full AM mirror. The geometry of the part \SI{200}{\milli\meter} (length) $\times$ \SI{100}{\milli\meter} (width) $\times$ \SI{10}{\milli\meter} (height) posed a challenge with the \SI{2}{\milli\meter} thick mirror surface plus concave curvature resulting in a high attenuation of X-rays, when the mirror was side-on (parallel) to the X-ray source/cone. This led to a poor contrast between the part and air, making it difficult to identify part edges and pores and thereby reducing the accuracy of pore quantification. Further, the dimensions of the mirror limited the voxel size to \SI{60}{\micro\meter} which posed a challenge as pores can be tens of micrometres in diameter and are only considered significant if they are represented by several voxels. The porosity detection resolution is often greater than this due to the XCT best practice of treating single-segmented voxels as noise. Based upon these findings, it was decided to take XCT scans of cut sections of the mirror prototypes.

Smaller sections allow the parts to be placed closer to the X-ray source, increasing the magnification on the detector and minimising the voxel size. This, in turn, enhances the porosity detection resolution, thereby improving the accuracy of porosity detection.  Reducing the aspect ratio of the part via sectioning increased the greyscale contrast, allowing for a more accurate identification between aluminium and air/pore. To select the appropriate dimensions for the sections of the mirror in this study, a virtual XCT simulation \cite{F.VidalgVXR} was used to assess the quality of the pixels at the different projections. CAD models of the final design were used for numerical simulations to predict the 3D representations of the mirrors formed by XCT. These simulations assumed the best-case scenario, that is zero porosity, as the porosity could not be predicted.

\subsection{Virtual XCT}
gVirtualXray (gVXR) was used to simulate X-ray projections around the sample~\cite{gVXR}. The X-ray source and detector specifications were taken from the XCT system specifications at KU Leuven, to obtain accurate simulations quickly\cite{POINTON2023107500,Vidal2024SPIE}. Corresponding XCT reconstructions were performed with Core Imaging Library (CIL)~\cite{doi:10.1098/rsta.2020.0192} and the code automatically optimised the geometric magnification. It reduced the source-to-object distance (SOD) so that the object appeared as big as possible in the X-ray projections without touching the source. Consequently, the pixel sizes of the projections were as small as possible, resulting in the identification of the smallest possible defects identifiable by the XCT scanner used.

\begin{table}[tbhp]
    \centering
    \caption{Optimised SOD for each sample.}
    \label{tab:SOD summary}
    \begin{tabular}{|c|c|c|c|c|}
\hline
       & Bounding box & SOD   & Pixel size    & \# of pixels below \\
Sample &  [mm]        &  [mm] &  [{\textmu}m] &  5\% of transmission\\
\hline
AsPrinted\_STL               & $79 \times 79 \times 13$ & 340.79 & 62.42  & 11\% \\ 
MirrorCut1\_MountSection     & $29 \times 21 \times 13$ & 97.49 & 17.77 & 0\% \\  
MirrorCut2\_Lattice+Surface1 & $38 \times 12 \times 13$ & 117.79 & 21.57 & 0\% \\  
MirrorCut3\_Lattice+Surface2 & $38 \times  6 \times  7$ & 113.70 & 20.83 & 0\% \\ 
\hline
\end{tabular}
\end{table}

\begin{figure}[b]
\centering
\includegraphics[width=0.85\linewidth]{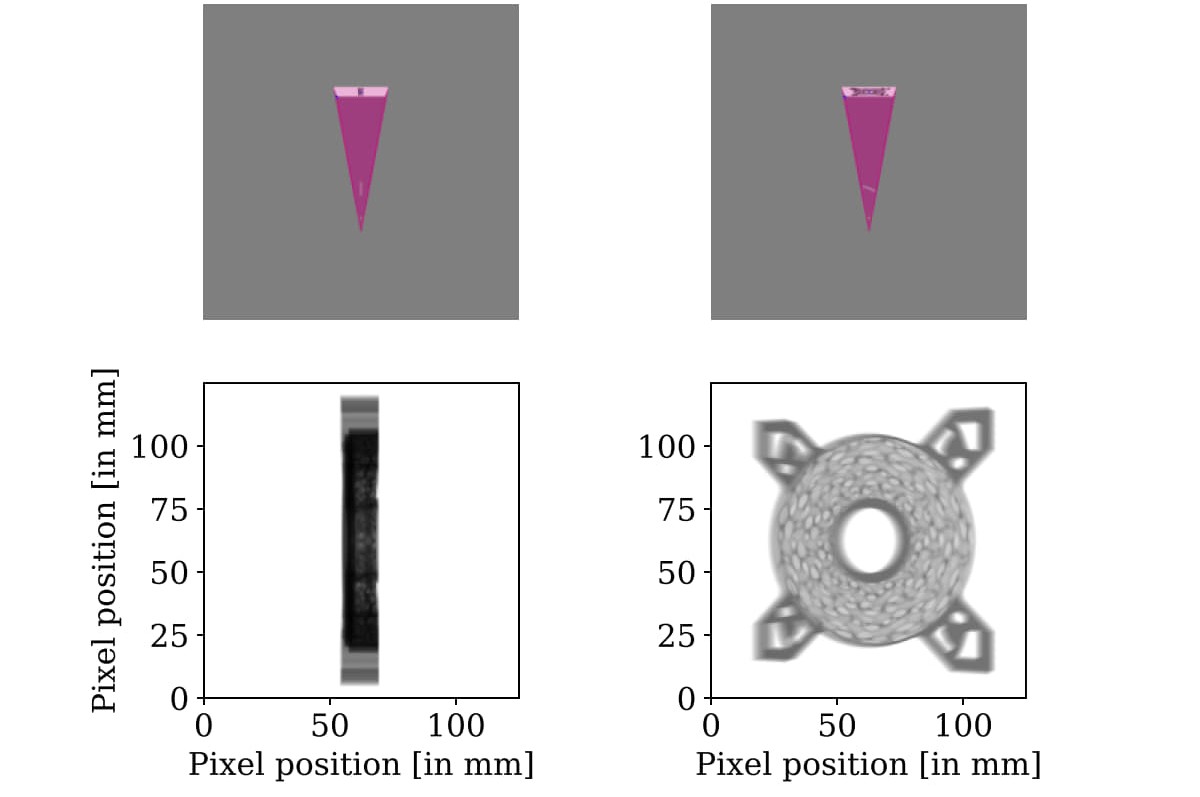}
\caption{Minimum and maximum coverage for the whole mirror with a SOD of \SI{340.79}{\milli\meter}.}
\label{fig:Whole mirror - coverage}
\end{figure}

\begin{figure}[tbhp]
\centering
\includegraphics[width=0.55\linewidth]{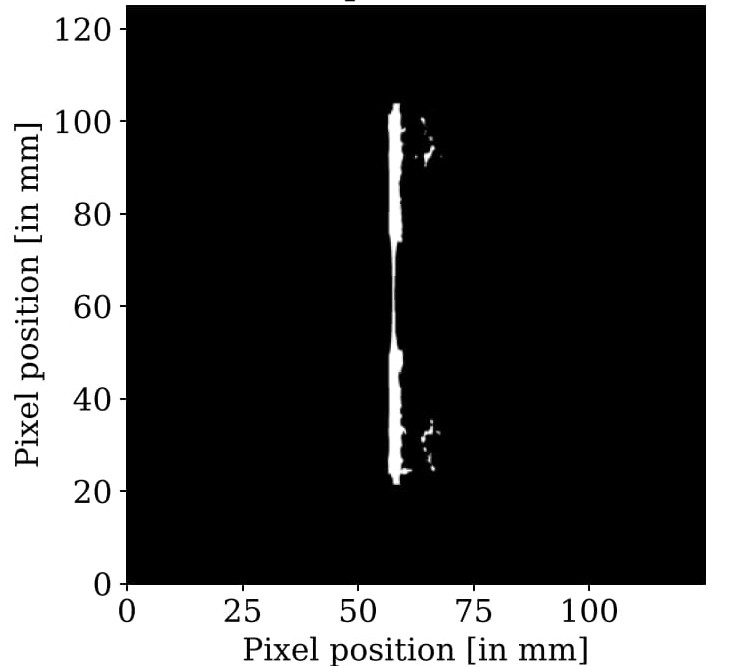}
\caption{Illustration of where photon starvation will occur in the lower left-hand side image of Figure~\ref{fig:Whole mirror - coverage}.}
\label{fig:Whole mirror - starvation}
\end{figure}

 A summary of the parameters is provided in Table~\ref{tab:SOD summary}. When the whole mirror is considered, the smallest pixel size achievable with the XCT device is 62.46 {\textmu}m. The orientation resulting in the greatest attenuation of X-rays, Figure~\ref{fig:Whole mirror - coverage} \textit{left}, was simulated to estimate if photon starvation will occur. In this projection, the maximum number of photons emitted by the X-ray source will be absorbed by the sample and will not reach the detector, and artefacts are likely to occur. When the whole mirror is considered, 11\% of the pixels have a photon transmission below 5\%, i.e., corresponding to photon starvation as shown in Figure~\ref{fig:Whole mirror - starvation}. The gVXR analysis highlighted photon starvation in scans of the mirror as a single part, which would result in a poor reconstruction. The same approach was applied to the proposed sections of the mirror, as shown in Table~\ref{tab:SOD summary}, which highlighted that the sections would not be subject to photon starvation and that the voxel sizes could be decreased. 

\subsection{XCT analysis}
\begin{figure}[htbp]
\begin{center}
\includegraphics[width=0.95\linewidth]{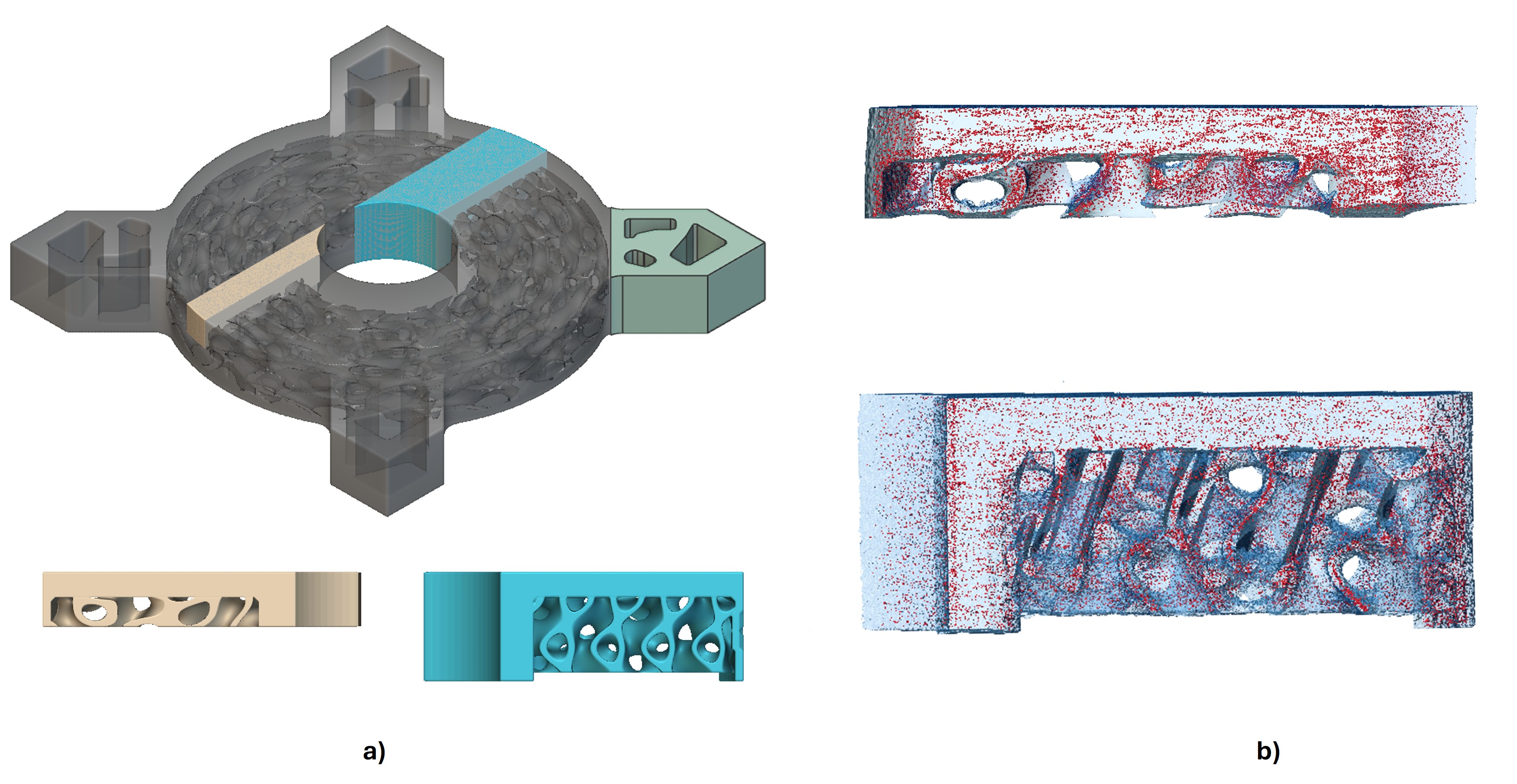}
\end{center}
\caption{Mirror cuts scanned for the XCT analysis: a) CAD file highlighting the three mirror cuts, Mirror Cut 1 Mount Section is coloured green; Mirror Cut 2 Lattice + Surface 1 is coloured blue; and Mirror Cut 3 Lattice + Surface 2 is coloured cream; b) porosity shown within Mirror Cuts 2 and 3.}
\label{fig:XCTAnalysisOverview}
\end{figure}

The wire EDM cut mirror sections scanned using XCT are shown in Figure~\ref{fig:XCTAnalysisOverview} a). A Nikon XTH 225 ST machine was employed with XCT scan settings as follows: \SI{150}{\kilo\volt}, \SI{54}{\micro\ampere}, \si{3142} projections with \SI{1000}{\milli\second} of exposure time for projection, and an achieved voxel size as reported in Table~\ref{tab:XCTvoxelsize_thresholding}. Avizo 2024.1 software was used to process and threshold the scan data. The XCT scans were first converted from 32-bit to 8-bit, then filtered using non-local means filtering, which reduced background noise and provided greater contrast between the pores and aluminium substrate. The pores were segmented using the interactive thresholding tool, which employs the global thresholding method, with the threshold values shown in Table~\ref{tab:XCTvoxelsize_thresholding}: voxel values above this threshold indicated air and values below, aluminium alloy. Pores of $\leq$ \SI{8}{voxels} were removed to filter out non-pore inclusions. The resultant processed scans with highlighted pores for Cuts 1 and 2 are shown in Figure~\ref{fig:XCTAnalysisOverview} a).

\begin{table}[htbp]
\caption{XCT analysis sample's voxel sizes and thresholding values.} 
\label{tab:XCTvoxelsize_thresholding}
\centering
\begin{tabular}{|l|l|l|}
\hline
 & {Voxel size [{\textmu}m]} &\begin{tabular}[c]{@{}l@{}}Part to background\\  thresholding value\end{tabular}\\ \hline
Cut 1 & 18 & 40 \\ \hline
Cut 2 & 21 & 76 \\ \hline
Cut 3 & 20 & 67 \\ \hline
\end{tabular}
\end{table}

\begin{figure}[htbp]
\begin{center}
\includegraphics[width=0.95\linewidth]{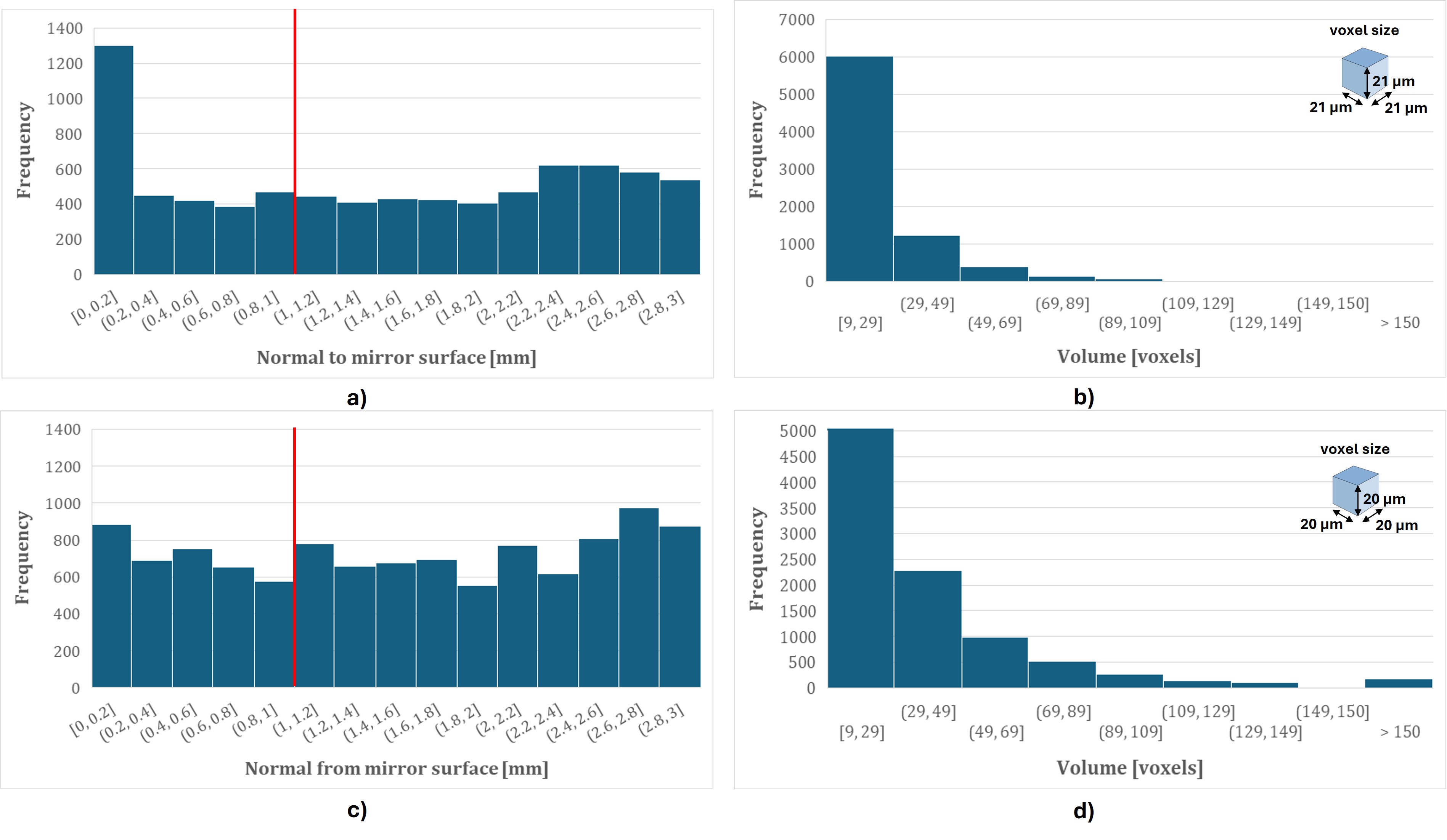}
\end{center}
\caption{Cut 2 and 3 XCT analysis data visualisation: a) Cut 2 plot of surface porosity by normal to surface, location of the mirror surface post machining highlighted in red; b) Cut 2 plot of porosity volume by frequency; c) Cut 3 plot of surface porosity by normal to surface, location of the mirror surface post machining highlighted in red; d) Cut 3 plot of porosity volume by frequency.}
\label{fig:XCTPlots}   
\end{figure}

Figure~\ref{fig:XCTPlots} a) visualises the frequency of pores by distance from the mirror surface, from this plot it can be seen Cut 2 had a significant cluster of pores within \SI{0,2}{\milli\meter} of the mirror surface with all remaining porosity distributed evenly down to the lattice structure. Figure~\ref{fig:XCTPlots} b) shows a plot of Cut 2's porosity volume by frequency. Figure~\ref{fig:XCTPlots} c) then visualises the frequency of pores by distance from the mirror surface for Cut 3 with an even distribution of pores throughout. Cut 3 also had a majority of pore volumes between \si{9} and \SI{30}{voxels} as shown in Figure~\ref{fig:XCTPlots} d). It is challenging to form comparisons between Cuts 2 and 3 due to the different thresholds used, which resulted in different attenuation of X-rays, also, Cut 3 is smaller than Cut 2, and so the latter would be expected to have fewer pores. However, conclusions can be drawn from this dataset, those being, first, that the majority of pores are small, although some of the identified pores are suspected to be noise. Second, that porosity was observed throughout the optical surface, highlighting the presence of pores after machining. Future work in this analysis will examine the sphericity of the pores, adjust threshold values to evaluate the impact on pore count, and consider sectioning future samples with a lower aspect ratio and smaller dimensions to reduce voxel size. 

\begin{figure}[htbp]
\begin{center}
\includegraphics[width=0.95\linewidth]{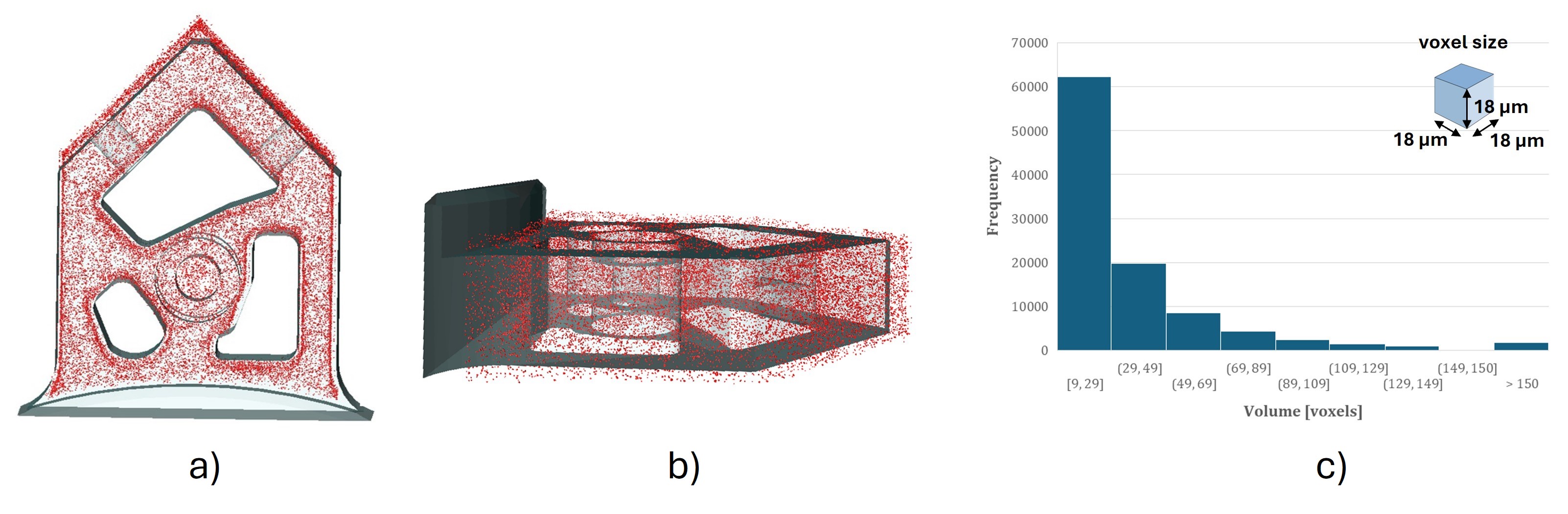}
\end{center}
\caption{Cut 1 XCT analysis data visualisation: a) top down view (faced down during printing) of the mounting structures porosity overlayed onto an as machined mounting structure; b) side on view of the mounting structures porosity overlayed onto an as machined mounting structure; c) Cut 1 plot of porosity volume by frequency.}
\label{fig:Cut3Analysis}
\end{figure}

One of the four mounting structures was also subject to XCT analysis, with the results shown in Figure~\ref{fig:Cut3Analysis} highlighting the location of pores relative to the machined surfaces. The higher density of pores along the perimeter, due to the laser turning back at these locations as shown in Figure~\ref{fig:printscanstrategy} a), could have affected the quality of the part after becoming exposed due to machining depending on the exact surface and how much if any material was removed. Specifically, the corner faces, which were machined flat for mounting and the side facing mounting holes. As a result, the mounting structure is likely to have a lower mechanical strength than simulated. A plot of pore volume by frequency was also constructed, with the same trend of the majority being the smallest pores. The lower threshold value for Mirror Cut 1 versus the other two cuts highlights the limitations in forming a comparison and reduces the confidence in the results, although, from Figure~\ref{fig:Cut3Analysis} a) it is clear that the laser path influences porosity.

A significant limitation of XCT is that the results obtained using this method are untraceable to a certified standard, making it difficult to quantify measurement errors accurately\cite{Chahid_2024}. One approach considered for confirming the accuracy of porosity identification was to use a small calibration pin with micro-drilled (\SI{100}{\micro\meter} \diameter) holes within it; however, despite using the calibrated holes for thresholding, confidence was low when applied to the bulk porosity, this approach and a discussion of alternative approaches, is provided by \textit{Chahid et al. (2024)} \cite{Chahid_2024}. A further limitation was selecting the optimum dimensions of the samples so that they are representative of the mirror, but also provided the highest confidence in the data, this is an ongoing effort which will utilise simulations to support dimension selection. Ultimately the XCT data can only be validated with external measurements. 

\section{Optical Metrology}
\label{sec:OpticalMetrology}
In preparation for environmental testing, a quantitive evaluation of the optical surface was required to form an understanding of the impact of environmental testing on the reflectivity of the mirrors. This was achieved through the identification and quantification of defects and areal images of the optical surface, where metrology was performed on the part using four external analysis techniques, which were as follows:

\begin{enumerate}

\item Optical microscopy - pores \& pore arrangement.

\item SEM - scratches, inclusions \& pores.

\item SEM - optical surface imaging.

\item Microscope interferometer - optical surface imaging.

\end{enumerate}

\subsection{Optical microscopy - pores \& pore arrangement}

\begin{figure}[htbp]
\begin{center}
\includegraphics[width=0.95\linewidth]{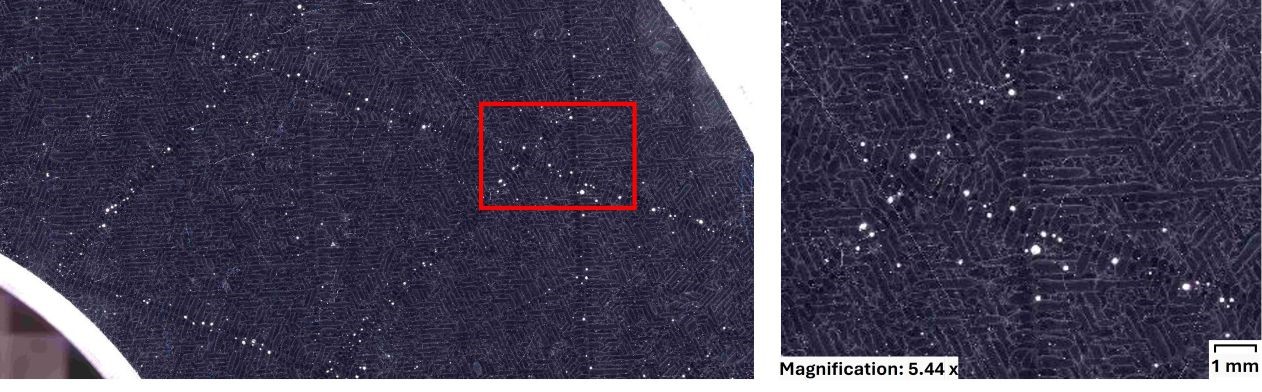}
\end{center}
\caption{Optical microscopy image of the Non-HIP Mirror \#5 using a darkfield observing mode; left: the clustering of porosity in a diamond pattern; right: a magnified region highlighting meltpools generated by the laser scan strategy.}
\label{fig:NonHIPHatchMarks}
\end{figure}

Microscopy images were taken with an Evident DSX2000 with a focal length of \SI{35}{\milli\meter}. The clustering of porosity in a diamond pattern seen on the non-HIP mirrors is clearly visible in Figure~\ref{fig:NonHIPHatchMarks}. This diamond pattern arises from the pores at the switchback points of the laser scanning strategy used during the print process, shown in Figure~\ref{fig:printscanstrategy}. The hatch lines are rotated \SI{67}{\degree} per layer; therefore, it is likely that the pores and optical surface transcend multiple layers. On-going efforts will quantify the number of pores per area and the diameter of the pores to compare against the XCT data and support the development of an optical scatter model.

\subsection{Scanning electron microscopy and Energy dispersive X-ray Spectroscopy}
SEM images were taken on a Zeiss Crossbeam 550 using three detectors. Two secondary electron detectors (InLens and SESI - secondary electron secondary ion) provided topography maps. The InLens detector, mounted normal to the measurement platform, provided the highest resolution but also flattened the image, whereas SESI, mounted at an angle with respect to the InLens, provided lower resolution but less flattened images. The third type of detector was an Energy Selective Backscatter (ESB) detector, which revealed the relative elemental composition of the target site through a greyscale representation; areas with a higher average atomic number appeared brighter, while lighter materials were depicted as darker. This contrast, known as atomic number (Z) contrast, showed that as the atomic number (Z) increased, so did the pixel intensity.  To provide an absolute measurement of elemental composition, energy dispersive X-ray spectroscopy (EDX/EDS) data was obtained using an Oxford Instruments XMax\textsuperscript{N} 150 with an accelerating voltage of \SI{5}{\kilo V} to \SI{20}{\kilo V}. This analysis determined the elemental composition of scratch initiators found on the mirror surface. An EDX analysis was performed at the initiation point of two observable scratch marks to identify their causes. Images were also taken of areas of interest, including surface inclusions, porosity and high magnification images to evaluate the differences between the HIP and non-HIP optical surface structures.

\subsubsection{SEM - scratches, inclusions \& pores}
\begin{figure}[htbp]
\begin{center}
\includegraphics[width=0.95\linewidth]{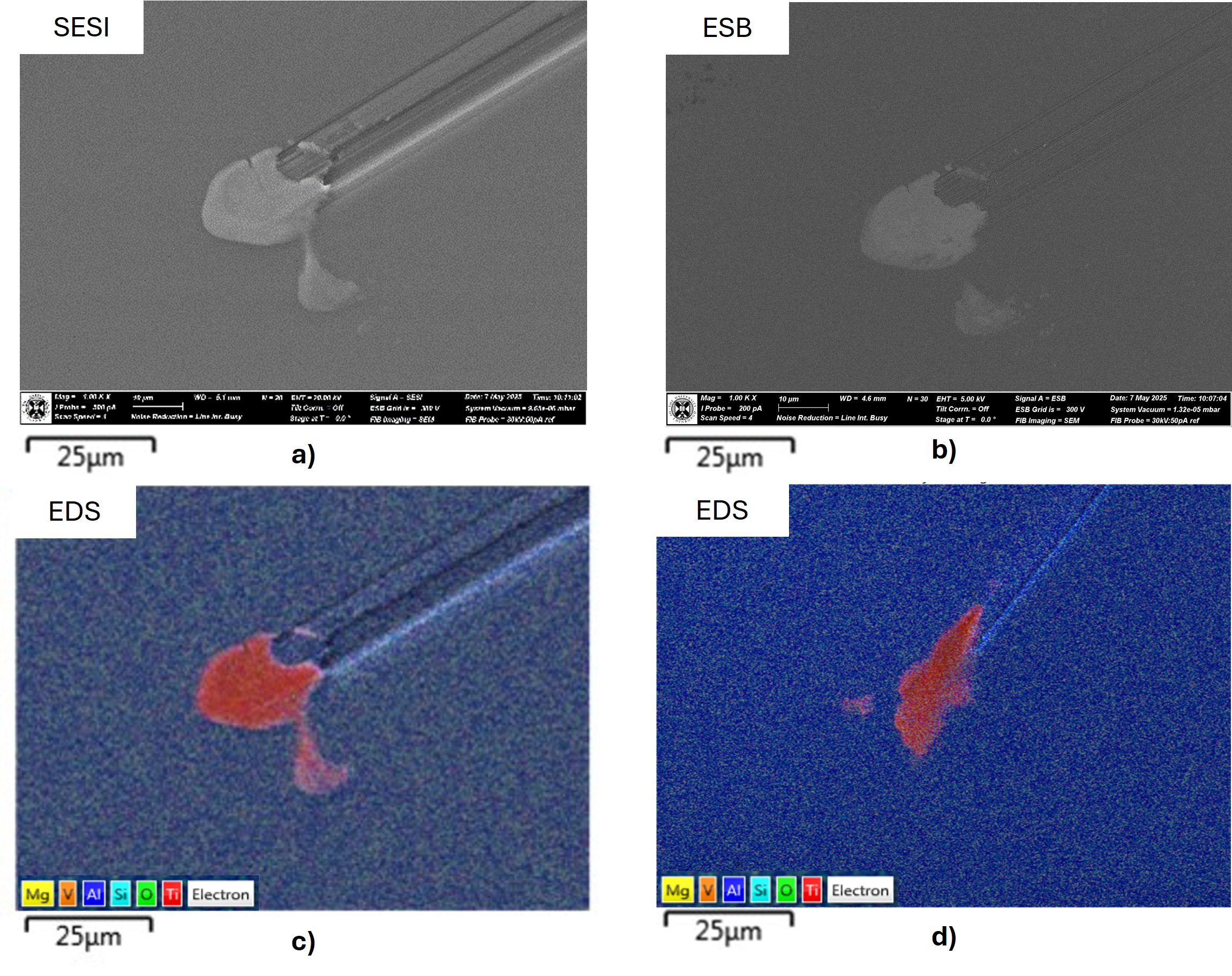}
\end{center}
\caption{SEM images of the mirror surface: a) SESI map of one of two scratch initiators observered on Mirror \#4; b) EBS image of the same scratch initiator as shown in a); c) EDS image of the same scratch initiator as shown in a) and b); d) EDS spectroscopy image of the second scratch initiator observed on Mirror \#4.}
\label{fig:SEMCollage}
\end{figure}

The two scratches observed on Mirror 4 were used to locate the inclusion that initiated the scratch. The SESI and ESB images shown in Figures~\ref{fig:SEMCollage} a) and b), respectively, suggest that the scratch initiator was atomically heavier than the AlSi10Mg alloy print material due to the lighter pixels. EDX scans of the scratch initiators shown in Figures~\ref{fig:SEMCollage} c) and d), determined that they were titanium vanadium alloy inclusions. This points to the TiAl6V4 titanium alloy, a commonly used AM powder, suggesting contamination within the print environment. This titanium alloy is harder than the aluminium alloy constituting the part and so would have scratched the surface during machining as described in the report by \textit{Wang et al. (2025)} \cite{wang2025effect}. Future work could include using XCT analysis to verify that the titanium inclusions originated from the print environment, with the titanium appearing as bright spots in the XCT scans.

\begin{figure}[htbp]
\begin{center}
\includegraphics[width=0.95\linewidth]{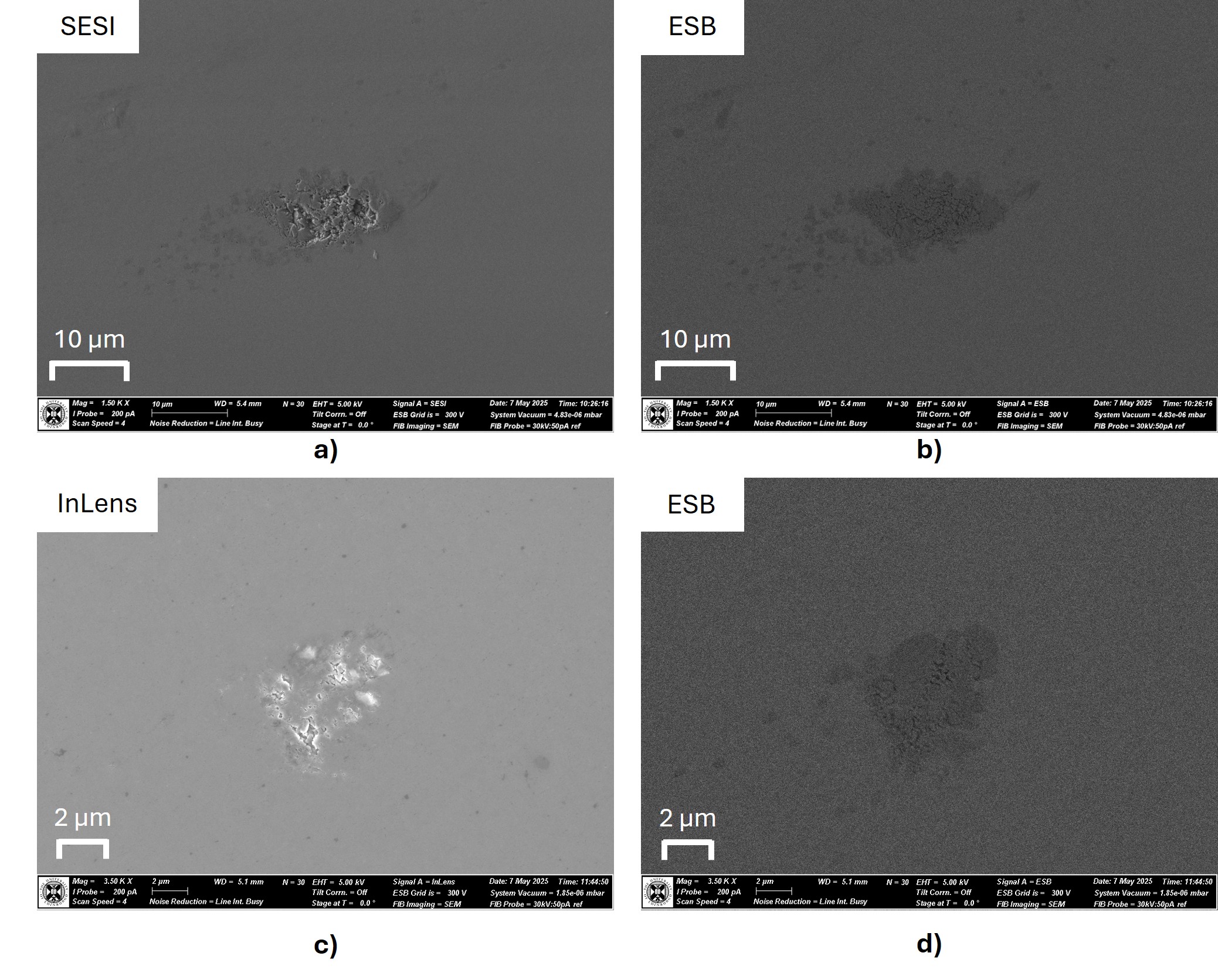}
\end{center}
\caption{SEM images of suspected oxide inclusions on the mirror surface: a) SESI map of a suspected oxide inclusion; b) ESB image of the same inclusion; c) InLens map of a different suspected oxide inclusion; d) ESB image of the same inclusion.}
\label{fig:oxidecollage}
\end{figure}

Suspected oxide inclusions, which are similar to those described by \textit{Raza et al. (2021)} \cite{raza2021degradation}, were also found, as shown in Figure~\ref{fig:oxidecollage}. Oxides are atomically lighter than the aluminium alloy constituting the part and therefore appear darker in the greyscale EBS images. However, these did not initiate any scratches, unlike the findings of \textit{Bai et al. (2020)} \cite{BAI2020116597} and \textit{Wang et al. (2015)} \cite{wang2015effect}, but as no EDX scan was taken, the elemental composition of these inclusions was not confirmed.

\subsubsection{SEM - optical surface imaging}
\begin{figure}[htbp]
\begin{center}
\includegraphics[width=0.9\linewidth]{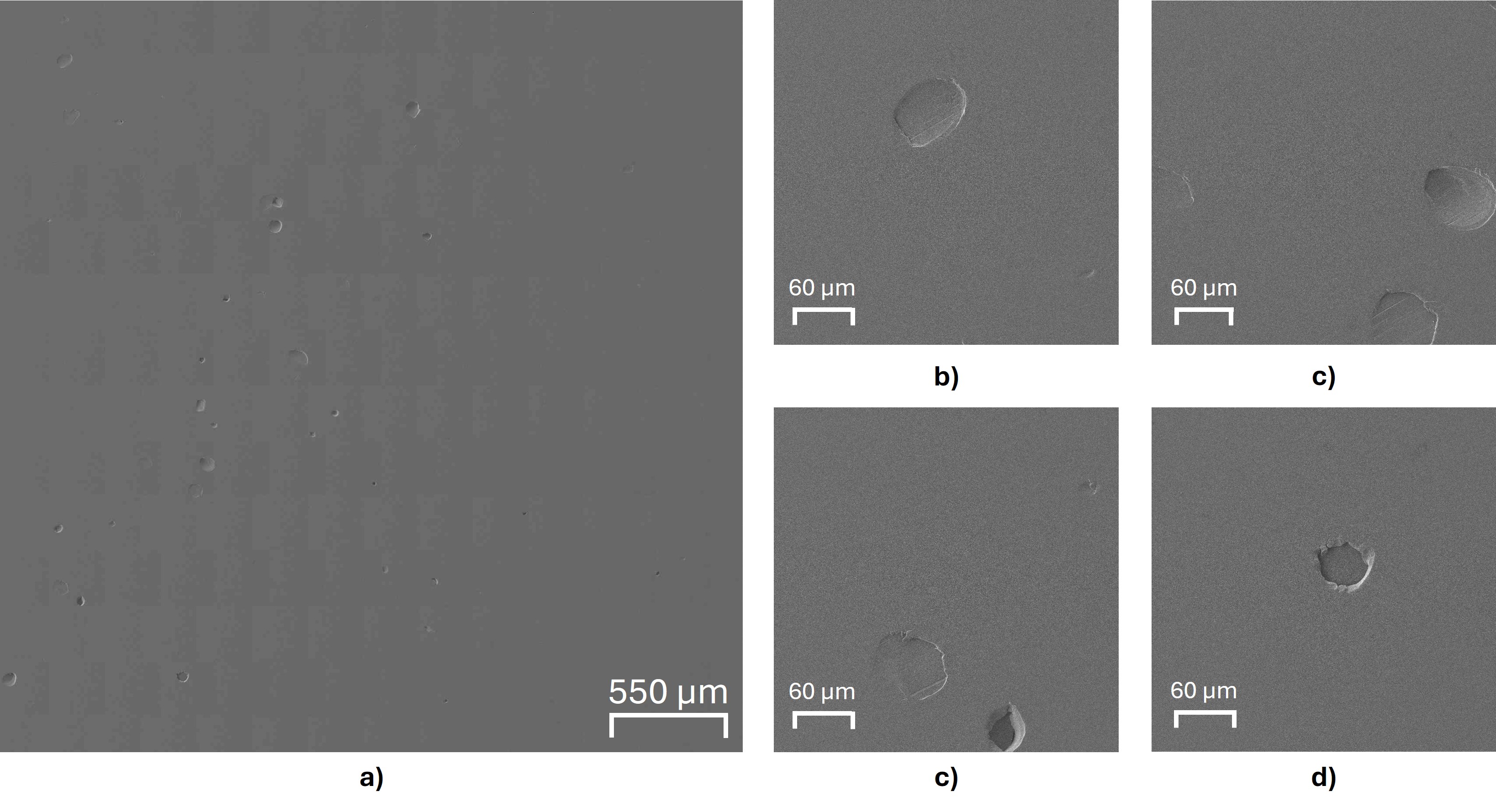}
\end{center}
\caption{SEM mapping of the surface porosity on non-HIP Mirror \#4: a) SESI mapping of the mirror surface porosity; b) to d) zoomed in images of individual pores seen in a).}
\label{fig:InLensSesiMap}
\end{figure}

To complement the optical microscopy and XCT data, surface pores on the mirror surface were imaged using large area mapping where multiple individual images were stitched together, as shown in Figure~\ref{fig:InLensSesiMap}. The scale used for these images was \SI{200}{\nano\meter} per pixel. From the SESI mapping (Figure~\ref{fig:InLensSesiMap}), the porosity mechanism can be inferred, as the pores are circular in shape they are suspected to be keyhole pores which form when the print conditions are too hot. This data will be further analysed in future to obtain more accurate information on pore size and correlated with the microscope and XCT data.

\begin{figure}[htbp]
\begin{center}
\includegraphics[width=0.9\linewidth]{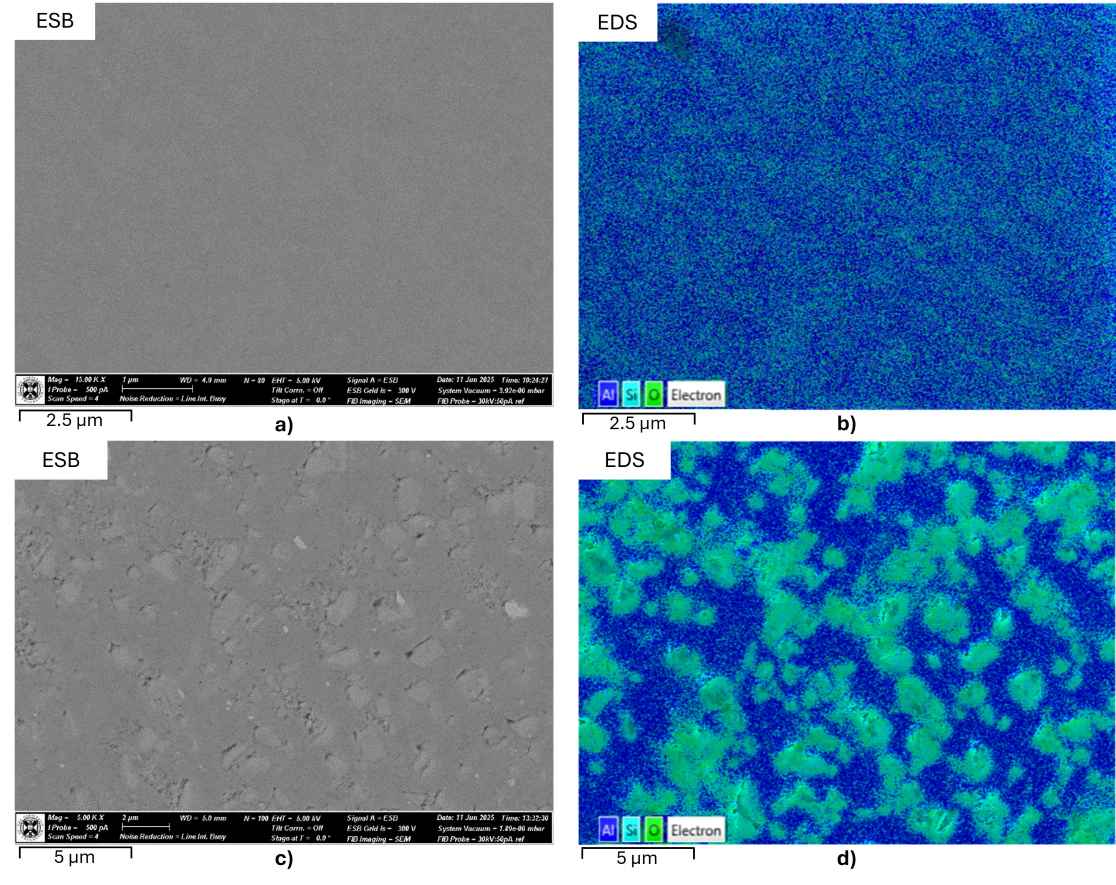}
\end{center}
\caption{High magnification SEM and EDX images of Mirror's \#2 and \#4: a) ESB image of Mirror \#4's non-HIP optical surface; b) EDS scan of Mirror \#4's non-HIP optical surface in the same region as shown in a); c) ESB image of Mirror \#2's HIP optical surface; d) EDS scan of Mirror \#2's HIP optical surface in the same region as shown in c).}
\label{fig:HighMagSurface}
\end{figure}

To examine the differences in microstructure between the HIP and non-HIP mirrors that lead to variations in reflectivity, higher magnification images of Mirrors \#2 and \#4 optical surfaces were taken for a quantitative analysis and comparison. A \SI{5}{\kilo V} accelerating voltage was used for these images to obtain a higher resolution in the EDX maps. Based on the ESB image and EDX scan shown in Figures~\ref{fig:HighMagSurface} a) and b), the surface of non-HIP mirror \#4 appears uniform in terms of the the Si and Al components.  The HIP surface, as shown Figure~\ref{fig:HighMagSurface} c), shows clear differences in the distribution of aluminium and silicon, with silicon precipitates identified in Figure~\ref{fig:HighMagSurface} d) via EDX scans. These silicon precipitates formed during the HIP process as explained by \textit{Wang et al. (2025)} \cite{wang2025effect}. It is suggested that the silicon precipitates then protrude from the surface post-machining by rebounding, resulting in increased surface roughness.

\subsection{Microscope interferometer - optical surface imaging}

\begin{figure}[htbp]
   \begin{center}
   \includegraphics[height=7cm]{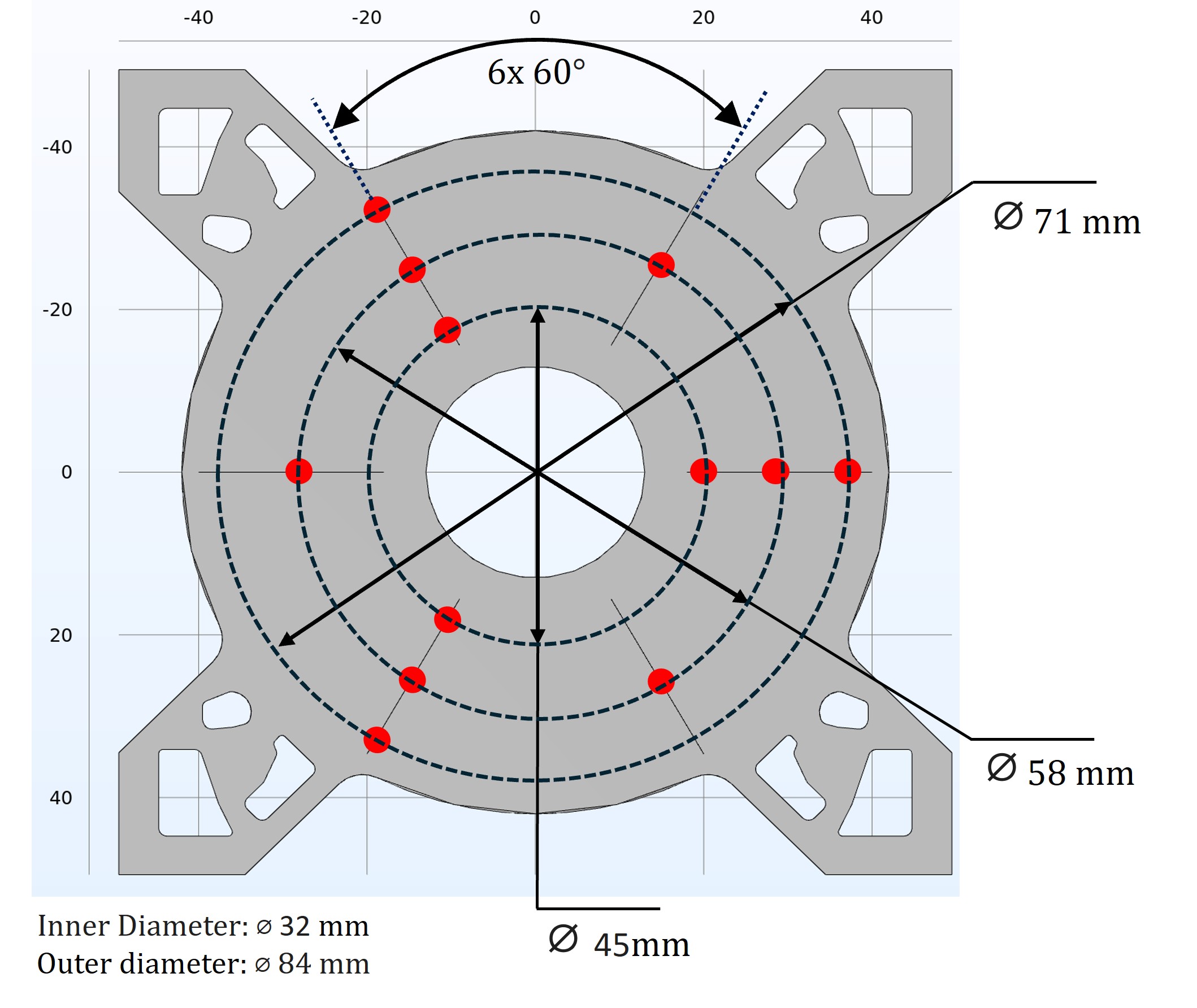}
   \end{center}
   \caption[example] 
   { \label{fig:surfaceroughness_measurementpoints} 
Diagram showing the location of the surface roughness measurement data points.}
   \end{figure}

Surface roughness measurements of the four SPDT processed mirrors were taken at Durham University using a Zegage scanning white light interferometer with a magnification of {$\times$}\SI{20} with a field of view of \SI{410}{\micro\meter} {$\times$} \SI{410}{\micro\meter}. An eighth-order polynomial has been subtracted from the data to remove form error contributions. 12 measurement points were evaluated and located along three circumferences of \SI{45}{\milli\meter}, \SI{58}{\milli\meter} and \SI{71}{\milli\meter} respectively with spacings of \SI{60}{\degree} between circumferentially adjacent measurement points, as shown in Figure~\ref{fig:surfaceroughness_measurementpoints}. This was to ensure that the data sampled the entire mirror surface while providing continuity across all four mirrors.

\begin{figure}[htbp]
\begin{center}
\includegraphics[width=0.95\linewidth]{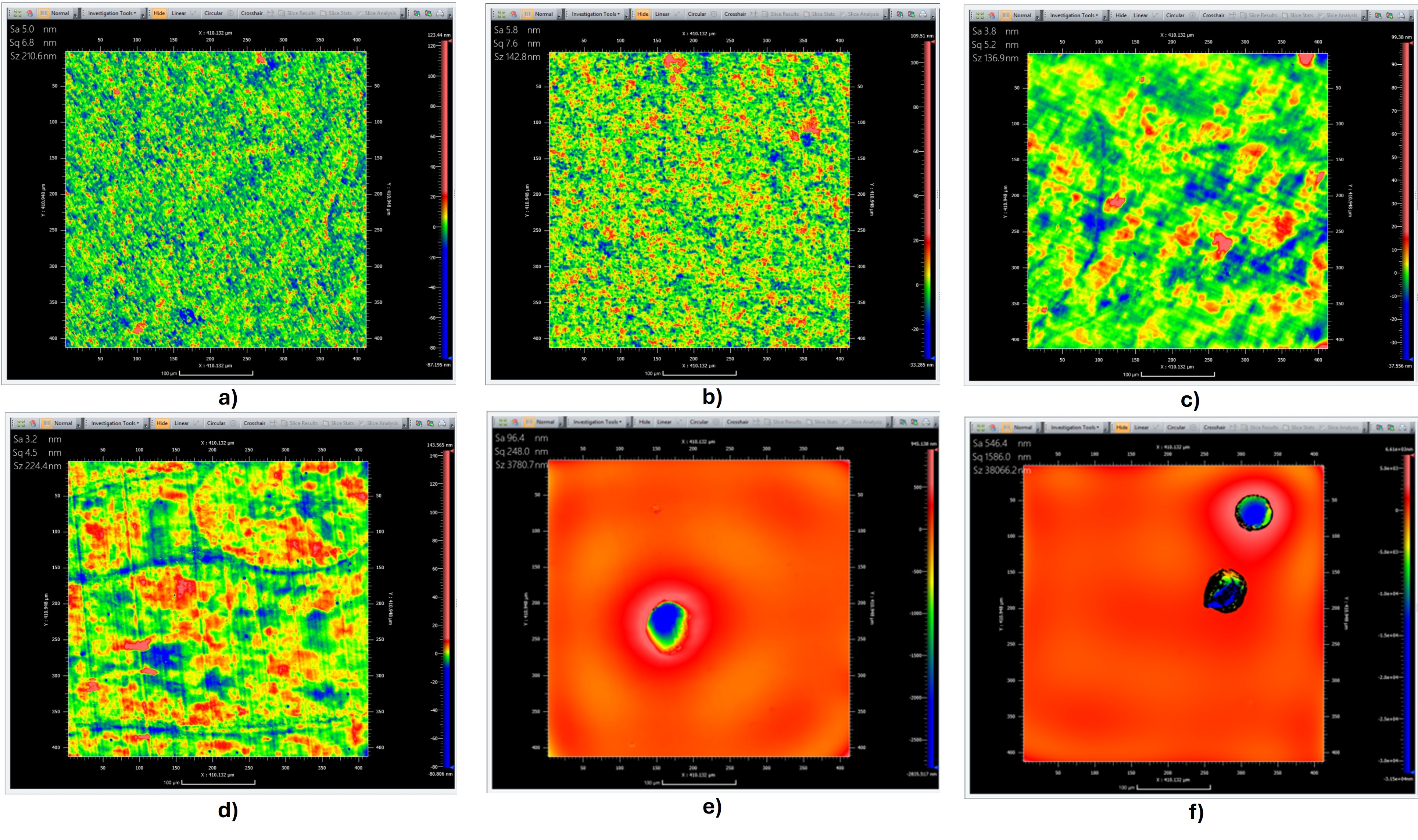}
\end{center}
\caption{Surface roughness microscope images of each of the four mirrors: a) Mirror \#2 data point 1; b) Mirror \#3 data point 5; c) Mirror \#4 data point 10; d) Mirror \#5 data point 12; e) Mirror \#4 artefact data point 5; f) Mirror \#5 artefact data point 2.}
\label{fig:surfaceroughness_collage}   
\end{figure}

Sa (mean value), Sq (RMS) and Sz (PV) values were obtained from the 12 measurement points across all four mirrors. As shown in Table~\ref{tab:SurfaceRoughnessRawData}, the two non-HIP mirrors had data points corresponding to defects which can be seen in Figures~\ref{fig:surfaceroughness_collage} e) and f), these localised defects created artefacts after polynomial subtraction. A statistical analysis was performed without the defect locations to obtain a best-case assessment of surface roughness. The prevalence of these defects is demonstrated in the SEM SESI map shown in Figure~\ref{fig:InLensSesiMap} a).

\begin{table}[htbp]
\caption{Surface roughness raw data to 2 significant figures, Sa (mean value), Sq (RMS), and Sz (PV) values; defect values are highlighted in red. Only mirrors \#2 and \#3 (shaded) were subject to HIP. An eighth-order polynomial has been subtracted from the data.} 
\label{tab:SurfaceRoughnessRawData}
    \centering
    \begin{tabular}{|p{1.8cm}||p{0.7cm}|p{0.7cm}|p{0.7cm}|p{0.7cm}||p{0.7cm}|p{0.7cm}|p{0.7cm}|p{0.7cm}||p{0.85cm}|p{0.85cm}|p{0.85cm}|p{0.85cm}|}
    \hline
         \multicolumn{1}{|l||}{} & \multicolumn{4}{c||}{\textbf{Sa {[}nm{]}}} & \multicolumn{4}{c||}{\textbf{Sq {[}nm{]}}} & \multicolumn{4}{c|}{\textbf{Sz {[}nm{]}}} \\  \hline
        \diagbox[width=2.2cm]{Pt \#}{Mirror} & \cellcolor[HTML]{E3F2FD} \#2 & \cellcolor[HTML]{E3F2FD} \#3 &  \#4  & \#5 & \cellcolor[HTML]{E3F2FD} \#2 & \cellcolor[HTML]{E3F2FD} \#3 & \#4 & \#5& \cellcolor[HTML]{E3F2FD} \#2 & \cellcolor[HTML]{E3F2FD} \#3 & \#4 & \#5  \\ 
        \hline
        1 & 5.0 & 6.1 & 4.4 & 3.0 & 6.8 & 7.9 & 5.9 & 3.8 & 210 & 160 & 220 & 170  \\ 
        2 & 5.2 & 5.7 & 3.7 & \cellcolor[HTML]{FFCDD2} 6.1 & 7.0 & 7.3 & 4.7 & \cellcolor[HTML]{FFCDD2} 33 & 130 & 150 & 130 & \cellcolor[HTML]{FFCDD2} 1800 \\ 
        3 & 5.0 & 5.9 & 4.0 & 3.1 & 6.6 & 7.5 & 5.7 & 4.4 & 180 & 97 & 160 & 220  \\ 
        4 & 4.9 & 5.9 & 3.5 & 3.2 & 6.5 & 7.5 & 4.4 & 4.9 & 100 & 95 & 58 & 98  \\ 
        5 & 4.9 & 5.8 & \cellcolor[HTML]{FFCDD2} 96 & 3.4 & 6.5 & 7.6 & \cellcolor[HTML]{FFCDD2}250 & 4.8 & 220 & 140 & \cellcolor[HTML]{FFCDD2} 3800 & 280  \\ 
        6 & 4.8 & 6.0 & 4.0 & 3.3 & 7.5 & 7.8 & 6.8 & 4.7 & 500 & 130 & 450 & 260  \\ 
        7 & 5.0 & 5.8 & 3.8 & 3.0 & 7.7 & 7.4 & 4.9 & 3.8 & 510 & 90 & 120 & 150  \\ 
        8 & 5.0 & 6.0 & 3.1 & 3.0 & 6.7 & 7.6 & 4.0 & 5.3 & 350 & 170 & 160 & 340  \\ 
        9 & 4.8 & 6.0 & 3.9 & 3.2 & 6.5 & 7.8 & 6.1 & 4.3 & 180 & 150 & 160 & 310  \\ 
        10 & 4.8 & 6.1 & 3.8 & \cellcolor[HTML]{FFCDD2} 550 & 6.4 & 7.8 & 5.2 & \cellcolor[HTML]{FFCDD2} 1600 & 160 & 130 & 140 & \cellcolor[HTML]{FFCDD2} -  \\ 
        11 & 4.9 & 5.9 & 4.0 & 3.2 & 6.5 & 7.6 & 5.4 & 4.7 & 110 & 140 & 170 & 350  \\ 
        12 & 4.7 & 6.0 & 3.6 & 3.2 & 6.3 & 7.7 & 4.6 & 4.5 & 73 & 120 & 82 & 220  \\ 
        \hline
        \multicolumn{13}{|l|}{}\\
        \multicolumn{13}{|l|}{Statistics - data highlighted in red excluded from the calculations}\\
        \hline
        Mean & 4.9 & 5.9 & 3.8 & 3.2 & 6.8 & 7.6 & 5.2 & 4.5 & 230 & 130 & 170 & 240 \\
        Median & 4.9 & 6.0 & 3.8 & 3.2 & 6.6 & 7.6 & 5.2 & 4.6 & 180 & 140 & 160 & 240 \\
        S. Dev. & 0.1 & 0.1 & 0.3 & 0.1 & 0.4 & 0.2 & 0.8 & 0.4 & 140 & 25 & 98 & 79 \\
        Min. & 4.7 & 5.7 & 3.1 & 3.0 & 6.3 & 7.3 & 4.0 & 3.8 & 73 & 90 & 58 & 98 \\
        Max. & 5.2 & 6.1 & 4.4 & 3.4 & 7.7 & 7.9 & 6.8 & 5.3 & 510 & 170 & 450 & 350 \\
        \hline
        \multicolumn{13}{|l|}{}\\
        \multicolumn{13}{|l|}{Average Sa, Sq and Sz for HIP (shaded) and Non-HIP}\\
        \hline
         \textbf{Mean} & \multicolumn{2}{|c|}{\cellcolor[HTML]{E3F2FD} 5.4} & \multicolumn{2}{|c||}{3.5} & \multicolumn{2}{|c|}{\cellcolor[HTML]{E3F2FD} 7.2} & \multicolumn{2}{|c||}{4.9} & \multicolumn{2}{|c|}{\cellcolor[HTML]{E3F2FD}180} & \multicolumn{2}{|c|}{200}\\
        \hline
    \end{tabular}
\end{table}

From Table~\ref{tab:SurfaceRoughnessRawData}, it is evident that the HIP samples have a greater surface roughness than the non-HIP mirrors. All of the mirrors demonstrated surface roughness results with $<$ \SI{8}{\nano\meter} RMS. The maximum Sq readings of the two HIP mirrors were greater than those of the non-HIP mirrors.  This agrees with the findings of \textit{Atkins et al. (2024)} \cite{AtkinsHIP2024} and \textit{Wang et al. (2025)} \cite{wang2025effect} who linked HIP to increased surface roughness. It is recognised that the non-HIP roughness disregards the effect of porosity, resulting in an incomplete understanding of the surface characteristics. This method was considered appropriate because incorporating porosity datasets that contain artefacts from form error correction could bias the data.

\subsubsection{Total integrated scatter}
\begin{figure}[htbp]
\begin{center}
\includegraphics[width=0.65\linewidth]{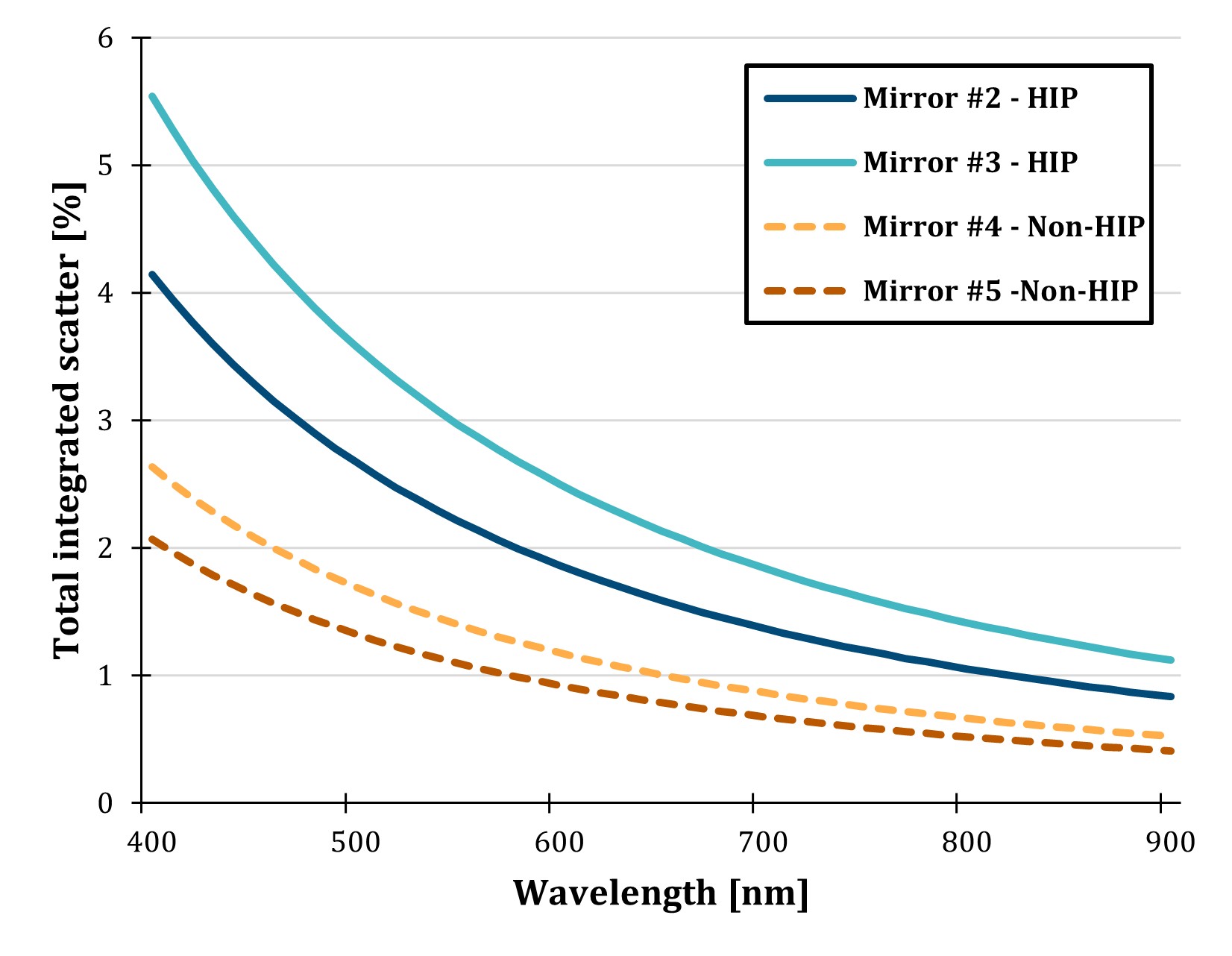}
\end{center}
\caption{Best case total integrated scatter (assuming no porosity) as a function of wavelength for each of the four mirror prototypes.}
\label{fig:TISPlot}
\end{figure}

\begin{equation}
    TIS = 1 - \exp\left(- (4 \pi \sin \theta \sigma / \lambda)^2 \right)
    \label{eq:TIS}
\end{equation}
An estimation of the total integrated scatter\cite{friedman2004photonics} (TIS) for the HIP and non-porous non-HIP mirrors can be calculated using Equation \ref{eq:TIS}. This equation assumes a Gaussian uniform distribution of roughness as a function of wavelength. Where \(\theta\) is the angle of incidence to the surface (\SI{90}{\degree}), \(\sigma\) is median RMS roughness (Sq) and \(\lambda\) is the wavelength of light. Due to the assumption of uniform surface roughness, the non-HIP TIS calculations assume no porosity is present; therefore, the calculation represents a best-case scenario and not the scatter anticipated from this mirror. The wavelength range chosen for this calculation was based on the CCam optical sensor specification \cite{CCamPaper}, the results of which are plotted in Figure~\ref{fig:TISPlot}. The HIP mirrors exhibit greater TIS as expected, with greater than 1\% scatter for the majority of wavelengths. The two non-HIP mirrors show suitability for infra-red (IR) wavelengths ($>$ \SI{750}{\nano\meter} with less than 1\% scatter). To calculate a more representative estimation of the scatter from the non-HIP mirrors and a more accurate comparison, the influence of porosity needs to be considered and this effort is described within the future work. 

\section{Summary and Future work}
\label{sec:SummaryandFutureWork}
This paper outlined the design, manufacture and evaluation of an AM aluminium alloy mirror which achieved a 56\% mass reduction compared to a solid body equivalent. A mirror design incorporating a split-p internal lattice structure with a field-driven ramp lattice thickness was selected as the final design. Five mirrors were printed in AlSi10Mg alloy using L-PBF. Two of these mirrors were subject to HIP and combined with an additional two underwent machining to achieve good optical surface quality with surface roughness $<$ \SI{8}{\nano\meter} RMS. XCT analysis determined the location and volume of porosity in the optical surface region as well as one of the mounting structures. External metrology of the optical surface identified titanium inclusion scratch initiators and linked HIP to a reduction in surface porosity, however, with increased surface roughness. 

\begin{figure}[htbp]
\begin{center}
\includegraphics[width = 0.95\textwidth]{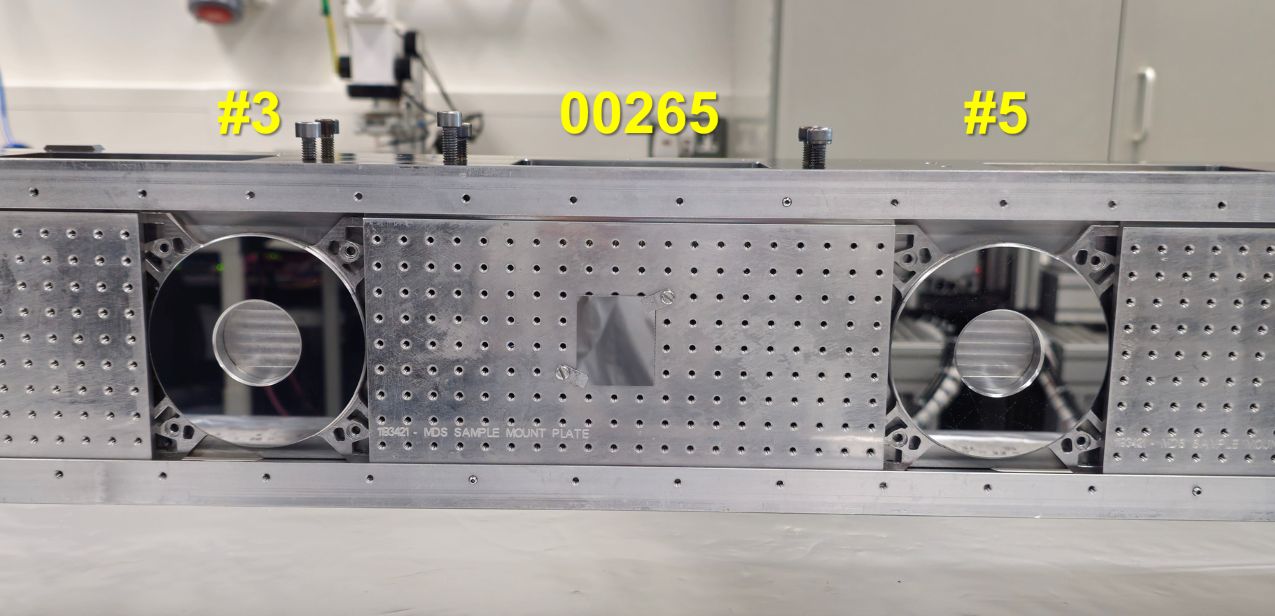}
\end{center}
\caption{Mirrors \#3 \& \#5 each with a \SI{50}{\nano\meter} chromium optical coating applied. Image credit: W. Tan and A. Majhi}
\label{fig:Final}
\end{figure}

Future work will include, firstly, the application of a \SI{50}{\nano\meter} chromium optical coating to HIP Mirror \#3 and non-HIP Mirror \#5, undertaken at the Diamond Light Source using a multilayer deposition system~\cite{HongchangWang}. The mirrors are displayed after the application of the optical coating in Figure~\ref{fig:Final}. Chromium was chosen for its ease of deposition and availability, rather than its reflectivity at the desired wavelengths; this choice was deemed suitable for initial characterisation and for starting environmental testing. Surface metrology will then be carried out to form a comparison with the non-optically coated mirrors. The mirrors will then undergo thermal cycle testing and further surface metrology to assess the effects of this testing on the optical surface quality.

In parallel, the external metrology data detailing pore size and surface roughness will be used to support the development of a bidirectional reflectance distribution function (BDRF) incorporating the scatter contribution from porosity. Through the development of a new model, a more accurate prediction of TIS can be calculated, which can inform long-term process development for AM mirrors. To support the validation of this model, experimental scatter measurements could be performed on Mirror \#4 (non-HIP).

\section{Data Availability Statement}

The data supporting this paper are openly available from eData, the STFC Research Data repository, at \href{https://edata.stfc.ac.uk/handle/edata/978}{edata.stfc.ac.uk/handle/edata/978}.

\section*{Acknowledgments}
The authors acknowledge the UKRI Future Leaders Fellowship ‘Printing the future of space telescopes’ under grant \# MR/T042230/1. Thanks are expressed to Martin Sanchez, Gregory Campbell, and John Horne for their support with the print process; to Sammuel Tammas-Williams for his input and advice on SEM imaging techniques; Hongchang Wang, Kawal Sawhney and Wadwan Singhapong for their guidance on optical coatings; Simon Alcock and Murilo Bazan Da Silva for conducting the post-coating surface roughness measurements; and Sean Leavey for supporting the optical microscope measurements.

\bibliography{report} 
\bibliographystyle{spiebib} 

\end{document}